\documentclass[lettersize,journal]{IEEEtran}
\usepackage{amsmath,amsfonts}
\usepackage{algorithmic}
\usepackage{algorithm}
\usepackage{array}
\usepackage{textcomp}
\usepackage{stfloats}
\usepackage{url}
\usepackage{verbatim}
\usepackage{graphicx}
\usepackage{cite}
\usepackage{subcaption}

\begin{document}

\title{Terahertz Chip-Scale  Meta-Networks with LSPR Routing: A Theoretical Framework}

\author{Maryam Khodadadi, ~\IEEEmembership{Member,~IEEE,} Hamidreza Taghvaee,~\IEEEmembership{Member,~IEEE,}  Pei Xiao, ~\IEEEmembership{Senior Member,~IEEE,} Gabriele Gradoni ~\IEEEmembership{Senior Member,~IEEE,} and Mohsen Khalily, ~\IEEEmembership{Senior Member,~IEEE,}
\thanks{Maryam~Khodadadi, Hamidreza~Taghvaee, Pei~Xiao, Gabriele~Gradoni~and~Mohsen~Khalily are with the Institute for Communication Systems (ICS), Home of the 5G and 6G Innovation Centre, University of Surrey, Guildford GU2 7XH, U.K. e-mails: m.khodadadi@surrey.ac.uk, and m.khalily@surrey.ac.uk.}
\thanks{The work of M. Khodadadi was supported by the Guarantee Funding for Horizon Europe Marie Skłodowska-Curie Actions (MSCA) Postdoctoral Fellowship, provided under Engineering and Physical Sciences Research Council (EPSRC) Grant No. EP/Z001552/1, United Kingdom.}}

\markboth{ }%
{Shell \MakeLowercase{\textit{et al.}}: A Sample Article Using IEEEtran.cls for IEEE Journals}


\maketitle

\begin{abstract}
Efficient chip-scale interconnects are critical for modern Microelectronic-photonic systems as they support high bandwidth utilization and low-latency processing. However, conventional wired links suffer from high resistivity and latency, while radio-frequency and millimeter-wave wireless solutions face limitations such as bandwidth congestion, interference, and power-inefficiency. Terahertz (THz) plasmonic communication, using surface plasmon polaritons (SPPs), enables high data rates and broad bandwidth for wireless network-on-chip (WiNoC) links, while remaining compatible with nanophotonic architectures. This paper presents a novel \textit{Binary Field-Driven Meta-Routing Method} supported by a semi-analytical framework that models the complex interaction between graphene’s tunable electromagnetic properties and THz plasmonic phenomena, by leveraging graphene impedance modulation to dynamically couple localized surface plasmon resonances (LSPRs) and guide them across a meta-network—enabling controlled beam steering within chip-scale architectures. The approach combines analytical conductivity models with coupled-mode theory and algorithmic control to predict and configure LSPR-based beam steering in graphene metasurfaces. Using this approach, four reconfigurable graphene meta-pixel antenna configurations---\textit{Y-MetaRouter}, \textit{MetaSwitcher}, \textit{Penta-MetaEmitter}, and \textit{CP-MetaCore}---were designed and analyzed. These configurations are capable of achieving unidirectional radiation, bi-directional meta-steering, frequency-driven multidirectional transitions, and circular polarization, respectively. Accordingly, real-time beam steering is enabled via chemical potential modulation, forming configurable LSPR pathways and creating virtual SPPs channels. 
To theoretically characterize these systems, a \textit{Coupled-Mode Theory of Field-Driven LSPR Meta-Networks} is formulated, modeling the current distribution of virtual SPPs and path-dependent LSPR coupling to predict far-field characteristics. Theoretical results show excellent agreement with full-wave numerical simulations. A point-to-point meta-wireless link is analyzed through theoretical and numerical methods, demonstrating scalability for low-latency, high-performance THz communication in WiNoC and nanophotonic platforms. System-level metrics—such as link budget, data rate, and reconfiguration energy—are estimated to validate feasibility for applications including chiplet communication, intra-core data transfer, heterogeneous computing, satellite systems, and compact transceivers in space-constrained environments. 
\end{abstract}


\section{Introduction}
\IEEEPARstart{T}{his} As modern processing architectures continue to scale in complexity and core count, efficient on-chip data communication has become a critical factor in sustaining overall system performance \cite{976921}. In many-core systems, the performance bottleneck is increasingly shifting from computation to interconnects, where traditional wired networks suffer from high latency, congestion, and energy inefficiency \cite{4120239,8289332}. These issues are further exacerbated by shrinking technology nodes, which increase wire resistivity and parasitic effects, leading to elevated power consumption and signal degradation \cite{906448}. As a result, the scalability of conventional interconnects is fundamentally constrained, particularly in bandwidth-intensive and latency-sensitive applications, underscoring the necessity of advanced Network-on-Chip (NoC) architectures as a scalable communication backbone for future computing platforms.

Wireless NoCs (WiNoCs) have emerged as a compelling alternative, offering low-latency and high-bandwidth communication through single-hop, contention-free links. Early efforts in this domain focused on radio frequency (RF) and millimeter-wave (mmWave) solutions, which demonstrated promising data rates within the radiative near-field \cite{4120239,8289332}. Adaptive and reconfigurable mmWave-based designs have further improved network performance and flexibility \cite{10529171,8999564}. However, these solutions face fundamental limitations, including bandwidth saturation, increased electromagnetic interference, and power-hungry multiple-input multiple-output (MIMO) architectures, making them less suitable for future many-core systems that demand higher scalability and energy efficiency \cite{MičicaWrightKolejákLezierPostavaHaweckerDeVetterTignonMangeneyJaffresLebrunTiercelinVanwolleghemDhillon+2024+1899+1907}.

Terahertz (THz) gap, spanning 0.1–10~THz, offers a transformative opportunity for on-chip wireless communication, promising ultra-high data rates, low latency, and enhanced spectral efficiency \cite{AKYILDIZ201416}. Nonetheless, THz integration introduces its own set of challenges. High propagation losses in semiconductor substrates, poor wave confinement, and limited diffraction control hinder practical deployment. Additionally, efficient and compact THz waveguides and antennas remain difficult to realize within dense chip environments. To address these limitations, researchers have explored leaky wave antennas (LWAs) for on-chip THz interconnects \cite{5871323,6518285}. Early LWA implementations based on complementary metal-oxide-semiconductor (CMOS) and metal-oxide-semiconductor field-effect transistor (MOSFET) technologies provided basic THz radiation capabilities \cite{6243783,6982195,7175082}, however they typically suffer from limited beam steering range, fixed radiation patterns, and low tunability—critical drawbacks in dynamic on-chip networks. In parallel, metasurfaces have emerged as a powerful solution to enable fine-grained wavefront control at subwavelength scales. These engineered surfaces manipulate electromagnetic (EM) waves through localized phase shifts, enabling functions such as beam steering, focusing, and signal redirection \cite{f2022metasurface,MOHSEN}.

\begin{figure*}
    \centering
    \includegraphics[width=\linewidth]{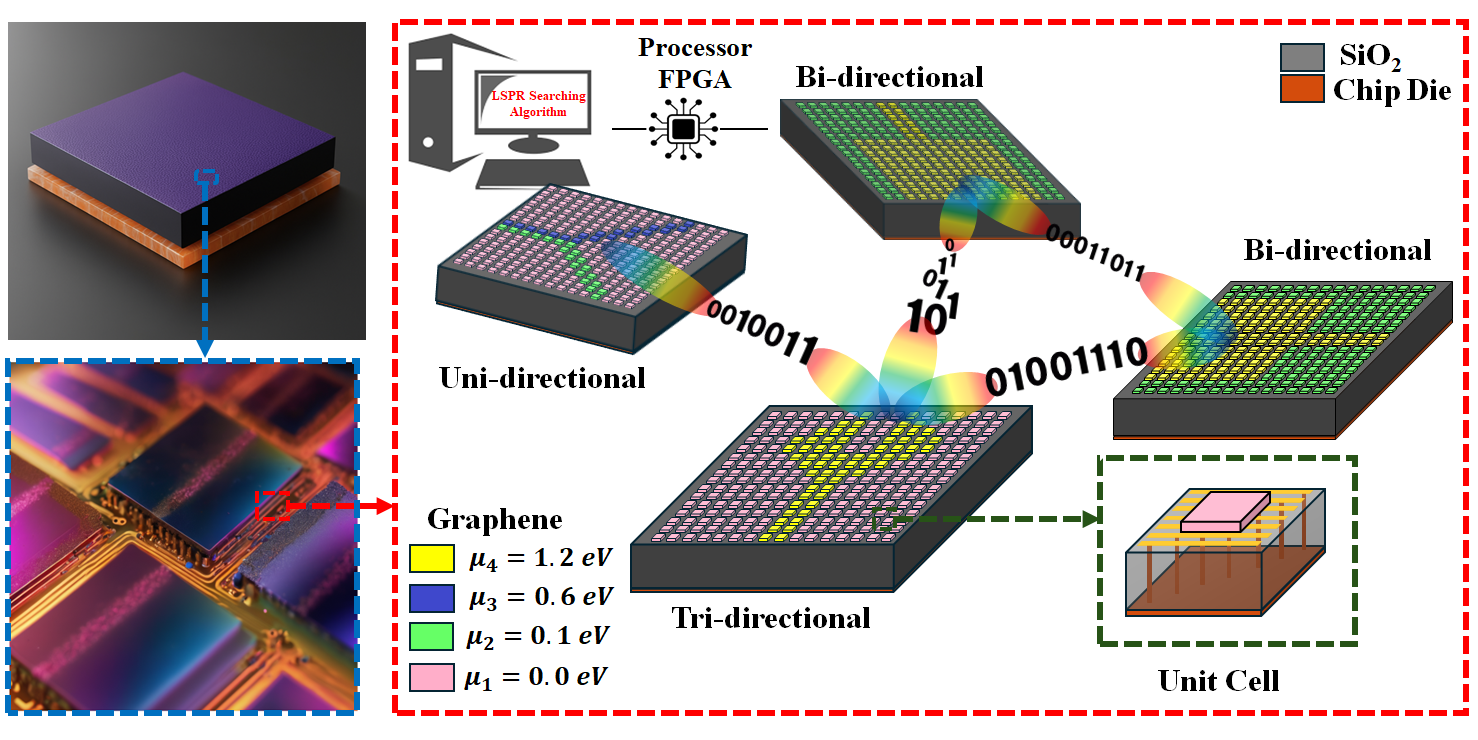}
    \caption{
Conceptual overview of the proposed graphene-based meta-pixel antenna platform for dynamic THz beam steering control. The meta-network consists of a programmable metasurface composed of graphene patches integrated on a SiO$_2$ substrate supported by a chip die. A processor (e.g., FPGA) executes the \textit{LSPR Meta-Routing Algorithm} to modulate the chemical potential ($\mu$) of individual meta-pixels, enabling selective excitation of LSPRs. By reconfiguring the activation pathways, various radiation modes are achieved—including uni-, bi-, and tri-directional beams—through real-time field-driven meta-routing. The digital modulation of $\mu_1 = 0.0$~eV to $\mu_4 = 1.2$~eV defines active and passive regions within the meta-network, supporting programmable THz wave propagation and multi-user communication. A 3D schematic of the unit cell is also shown for structural reference.
}
    \label{fig:fig1}
\end{figure*}

The advent of programmable metasurfaces marked a turning point in the field. Introduced by Cui et al. in 2014 \cite{Cui2014}, these structures bridge EM wave control with digital programmability, enabling dynamic, reconfigurable operations. Programmable metasurfaces offer unprecedented control over the phase, amplitude, and polarization of incident waves, and have demonstrated key functionalities for THz applications such as high-resolution focusing and agile beam steering \cite{10041955,PhysRevApplied.19.014004,9516921,PhysRevB.104.235409}. Despite these advances, implementation at THz frequencies remains hindered by the limitations of conventional tuning components—such as positive-intrinsic-negativ (PIN) and varactor diodes—which suffer from significant losses and reduced effectiveness at higher frequencies \cite{Zhang2018}.

To overcome these limitations, tunable materials such as liquid crystals \cite{0898}, barium strontium titanate (BST) \cite{4863669}, vanadium dioxide (VO$_2$) \cite{AmneElahi2020SIWCN}, and microelectromechanical systems (MEMS) \cite{5018865} have been investigated. Among these, graphene stands out due to its exceptional electronic and optical properties, including high carrier mobility, broadband tunability via electrostatic gating, and compatibility with CMOS processes \cite{KHAOUANI2021165429}. Graphene, a two-dimensional (2D) sheet of carbon atoms arranged in a hexagonal lattice, exhibits Drude-like conductivity at terahertz (THz) frequencies, making it highly suitable for active THz plasmonic components. Its high electron mobility and tunable conductivity support the excitation of highly confined, low-loss surface plasmon polaritons (SPPs), surpassing noble metals in the THz regime \cite{Barnes2003,ThanopulosKaranikolasPaspalakis+2024+4545+4554}. These properties are further enhanced by graphene’s sensitivity to external perturbations such as electrostatic or magnetostatic biasing, enabling dynamic control over its plasmonic response. A particularly promising approach involves the excitation of SPPs at the graphene–dielectric interface, which achieves extreme field confinement and efficient energy transport. These capabilities are pivotal for developing compact, high-performance THz metasurfaces with dynamic beam manipulation. Recent studies have demonstrated graphene-based metasurfaces capable of modulating beam direction and polarization in the THz domain \cite{Deng2017,Zhang2017,Wang2015}, laying the groundwork for reconfigurable wireless communication platforms.

Despite recent progress in graphene-based metasurfaces for THz communication, the realization of fast, fully dynamic reconfigurability remains elusive, limiting practical implementation of real-time beam steering and simultaneous scalable multi-user communication access. These limitations are particularly restrictive for NoC systems and emerging sixth-generation (6G) wireless architectures, where dynamic, scalable, and high-throughput communication is essential. Existing metasurfaces often rely on static or slow tuning mechanisms, resulting in high insertion losses and limited responsiveness in rapidly changing environments. While graphene-induced metasurfaces have demonstrated beam steering via oblique incidence and generalized Snell’s law~\cite{014001,1016}, real-time and simultaneous control of both phase and frequency remains largely unexplored at THz frequencies. However, beam steering alone is insufficient for densely integrated systems, where minimizing propagation loss between closely spaced devices is critical. Although the THz band offers vast bandwidth for high-speed communication, it also demands efficient, reconfigurable mechanisms to support multiple concurrent wireless links, especially in dynamic NoC and multi-user environments. As the number of interconnected devices—such as internet of thing (IoT) nodes and chiplets—increases, developing reconfigurable and tunable metasurface platforms becomes vital to ensure low-latency, energy-efficient, and adaptive communication.

This work bridges the gap between plasmonics and THz communication by investigating dynamically tunable graphene meta-pixel arrays capable of exciting localized surface plasmon resonances (LSPRs) for efficient THz wave manipulation. An innovative field-driven meta-routing algorithm is introduced, enabling the excitation of virtual SPPs and facilitating real-time beam steering, adaptive radiation pattern control, and polarization reconfiguration. This approach supports concurrent multi-user and programmable multi-beam routing communication, paves the way for scalable THz interconnects in future on-chip and chip-to-chip plasmonic systems. To validate the feasibility of the proposed meta-routing strategy, a generalized coupled-mode theory is formulated to model LSPR excitation, propagation, and emission into the chip domain, enabling accurate prediction of far-field characteristics. Finally, the scalability of the proposed method is validated through theoretical and numerical investigation of a point-to-point meta-wireless link, demonstrating its potential for low-latency, high-performance applications ranging from chiplet communication and many-core data transfer to satellite systems, advanced telecommunications, and compact THz transceivers.
\begin{figure*}
\centering
         \begin{subfigure}[b]{0.48\textwidth}
         \centering
         \includegraphics[width=\textwidth]{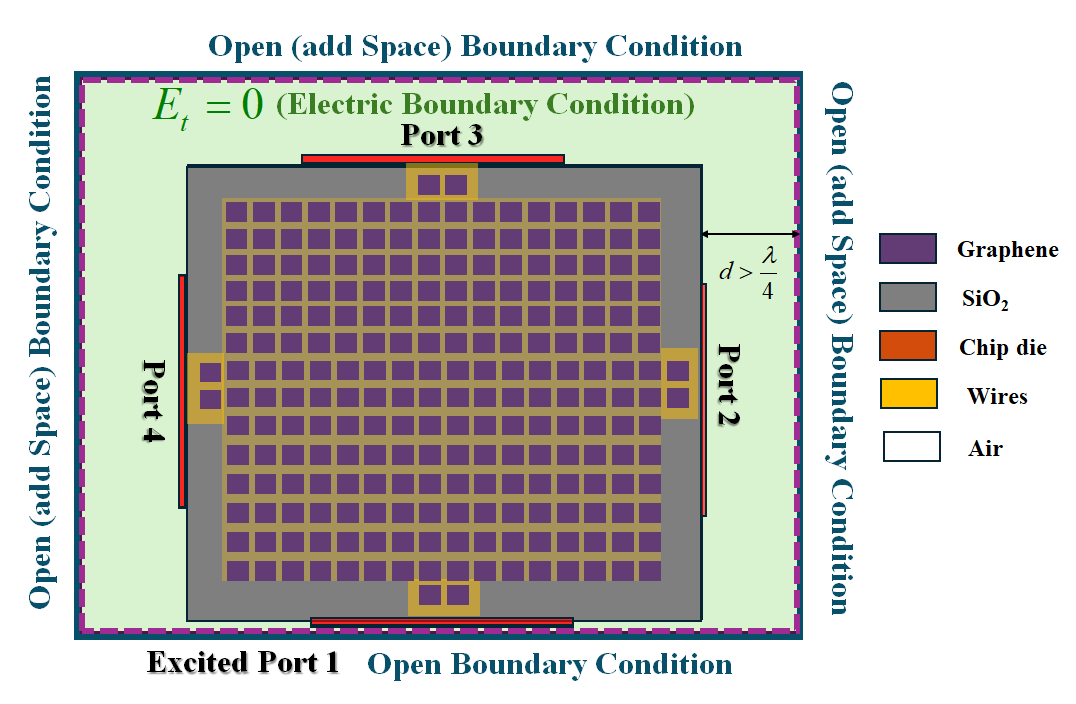}
         \caption{}
         \label{fig:figure2(a)}
     \end{subfigure}
              \begin{subfigure}[b]{0.48\textwidth}
         \centering
         \includegraphics[width=\textwidth]{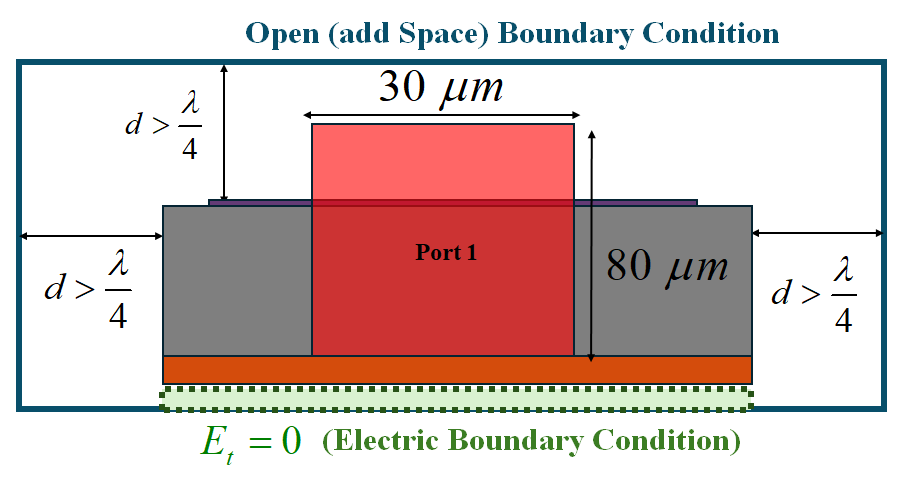}
         \caption{}
         \label{fig:figure2(b)}
     \end{subfigure}
    \caption{Numerical setup for full-wave simulation of network-driven excitation. (a) 2D top and (b) Side-view schematic illustrating boundary conditions and excitation port positioning for the metasurface simulation. \textbf{Note: The geometry is not drawn to scale; layer thicknesses and feature dimensions are schematically represented to illustrate structural relationships and boundary conditions.}}
    \label{fig:fig2}
\end{figure*}
The paper is organized as follows. In Section~II, the meta-pixel graphene configuration is introduced, along with the excitation and electrical biasing strategy used to control the antenna behavior. Section~III provides a detailed theoretical analysis of the plasmonic characteristics of the proposed meta-network structure. In Section~IV, the \textit{Field-Driven Meta-Routing Method} is presented, utilizing binary control over graphene meta-pixels to dynamically shape the LSPR coupling pathways. In Section~V, the \textit{Coupled-Mode Theory of Field-Driven LSPR Meta-Networks} is formulated to model the current distribution, impedance matching, far-field characteristics, and polarization behavior of the proposed configurations. In Section~VI, both theoretical and full-wave simulation results are presented for the reconfigurable graphene meta-pixel antenna designs—\textit{Y-MetaRouter}, \textit{MetaSwitcher}, \textit{CP-MetaCore}, and \textit{Penta-MetaEmitter}—demonstrating their ability to support unidirectional and multidirectional radiation patterns, as well as circular and linear polarization with real-time beam steering. In Section~VII, system-level performance metrics are estimated for a point-to-point on-chip THz communication scenario, using analytical link budget models and numerical validation to assess data rate, beam alignment tolerance, power efficiency, and switching latency. Finally, Section~VIII concludes the paper with a summary of the findings and a discussion on potential directions for future research in reconfigurable THz metasurfaces and WiNoC systems.

\section{On-Chip THz LSPR Pixel Architectures for Concurrent Directional Communication
}
As illustrated in Figure~\ref{fig:fig1}, the proposed on-chip LSPR meta-pixel architecture enables real-time programmable multi-beam routing through directional control of LSPR pathways in the THz regime. Different structural configurations are employed to achieve distinct radiation patterns—uni-directional, bi-directional, and tri-directional—by coupling the excited LSPRs in each unit cell to adjacent cells. This inter-cell coupling facilitates dynamic THz beam routing, which underpins the system’s multi-user capability.

The architecture consists of a $16 \times 16$ array of monolayer graphene patches, each $14$~µm wide, integrated onto a $SiO_2$ substrate with a relative permittivity of $\epsilon_{sub} = 2.09$. The substrate is $260$~µm thick and supported by a $40$~µm-thick perfect electric conductor (PEC) ground plane, which enhances THz field confinement and reduces radiation leakage. Each unit cell comprises a graphene patch patterned on the SiO$_2$ surface and separated from the ground plane, forming a tunable plasmonic resonator. By adjusting the chemical potential of the graphene in each meta-pixel unit cell (from $\mu_1 = 0.0$~eV to $\mu_4 = 1.2$~eV), the emission direction and intensity are controlled. A processor or field-programmable gate array (FPGA) executes an \textit{Binary Field-Driven
Meta-Routing Method}, See Section 4, to adaptively tune the array for real-time signal management and directional THz transmission.
In addition, the design exploits graphene’s ability to support slow waves, where the phase velocity is significantly lower than the speed of light in free space, enhancing wave–matter interaction and allowing subwavelength meta-atoms to effectively manipulate the THz waves. As a result, the meta-atoms can be much smaller and placed closer together (down to $\lambda/5$), making the whole structure more compact and efficient.
\begin{figure*}
    \centering
    \begin{subfigure}{0.49\textwidth}
    \includegraphics[width=1\textwidth]{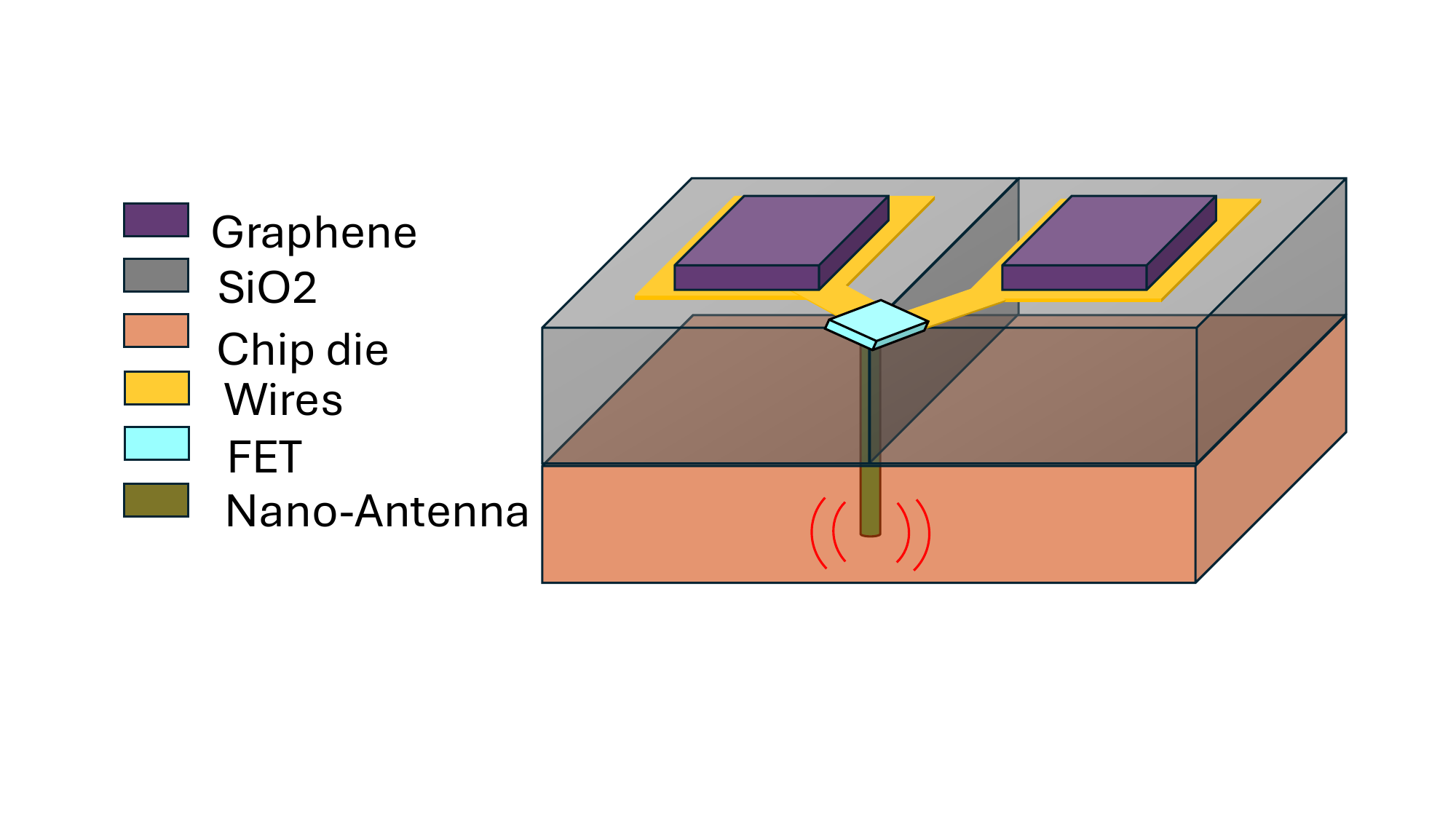}
             \caption{}
    \end{subfigure}
    \begin{subfigure}{0.5\textwidth}
    \includegraphics[width=1\textwidth]{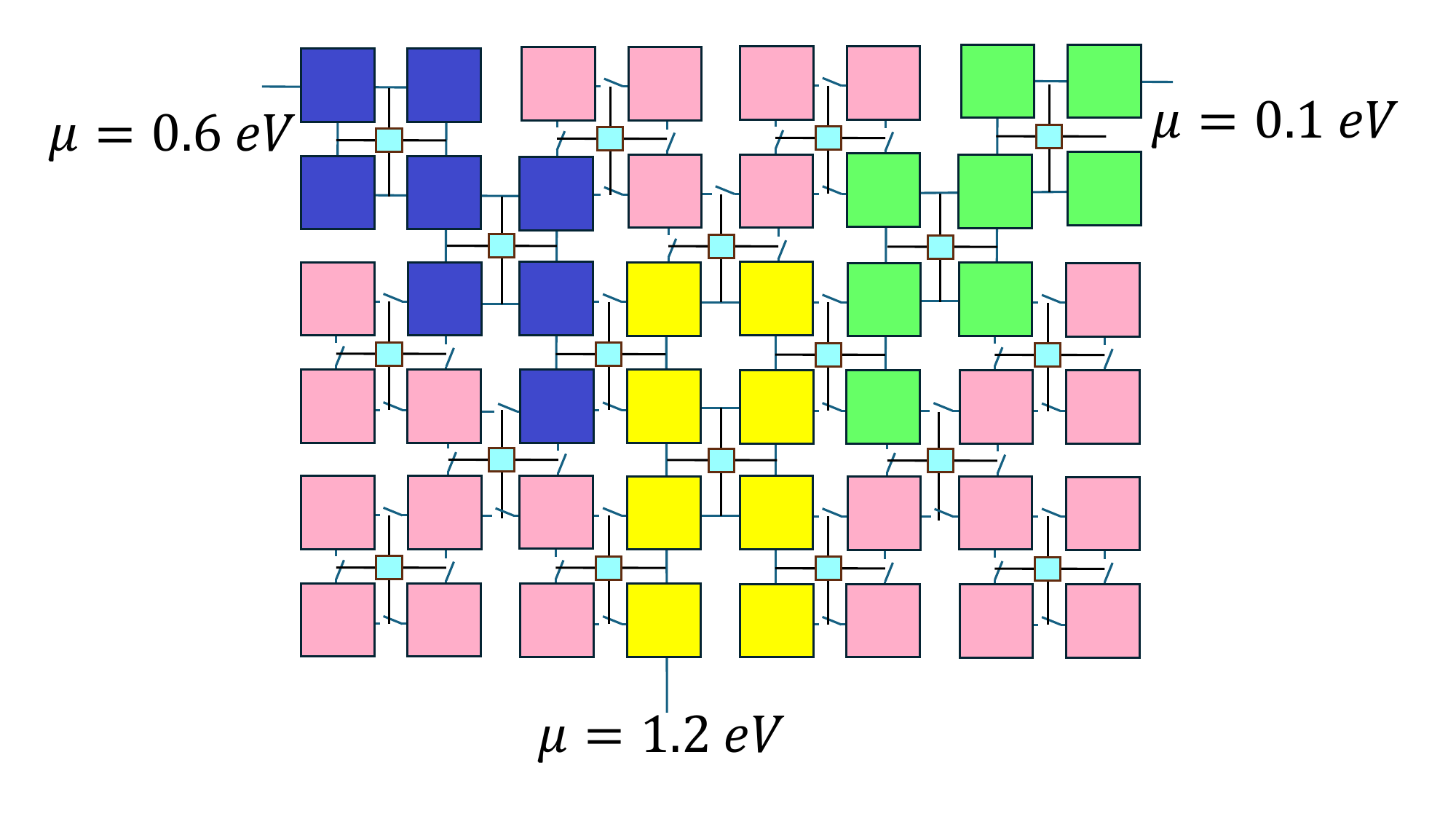}
             \caption{}
    \end{subfigure}
    \caption{Proposed biasing technique for tunable control of graphene’s chemical potential. (a) Architecture of a meta-pixel unit cell with integrated FETs and nanoantenna: A nanoantenna, integrated into the system, receives control signals that dictate the switching states of the FETs, creating a biased route of graphene pixels. (b) Feeding network: This schematic depicts a feeding network where multiple graphene meta-pixels are selectively biased through the use of FET switches. Each FET controls the connection of a specific route from the pixels to the voltage source.}
    \label{fig:fig3}
\end{figure*}
\subsection{Network-Driven Excitation and Biasing of Meta-Pixels}

As illustrated in Figure~\ref{fig:fig2}, the 2D LSPR-supporting meta-pixel array incorporates waveguide ports at the center of each side to enable programmable multi-beam routing and selective mode excitation at THz frequencies. These ports provide directional excitation, ensuring accurate modal excitation and strong mode confinement while suppressing higher-order modes. This design enhances both simulation fidelity and computational efficiency. To further reduce computational load, we impose a  PEC boundary at $z=0$ to model the ground plane and apply open boundary conditions at the waveguide ports to simulate active excitation. The remaining sides are terminated with open add-space conditions to minimize undesired reflections and support accurate field interaction modeling.

Figure~\ref{fig:fig3} introduces the proposed biasing and feeding network for the graphene meta-pixels. The meta-pixels are grouped into biasing domains, each controlled by field-effect transistors (FETs) acting as tunable switches. Biasing voltages are routed through peripheral patches, forming a network that facilitates plasmonic coupling across selected meta-pixels to enable dynamic LSPR routing. While the structure does not support arbitrary biasing of each individual patch, the designed wiring enables localized control within biasing domains, offering a practical compromise between tunability and circuit complexity.

To address potential interference from simultaneous wireless FET control signals, we propose time-division multiplexing or code-division schemes to isolate signal channels and prevent biasing conflicts. The biasing architecture reflects a balance between flexibility and hardware feasibility, enabling controlled manipulation of wave propagation across the metasurface. For command delivery, we build on intercell communication strategies as outlined in~\cite{8788546}, which support coordinated, reconfigurable voltage routing with minimal signal overhead.

\section{Theoretical Analysis of THz Plasmonics in 2D Meta-Pixel Arrays}
The interaction of THz waves with graphene can be described through its complex surface conductivity tensor $(\sigma(\omega,\mu,\tau,T,B_0))$, incorporating both diagonal and off-diagonal components~\cite{KHODADADI2022443}:
\begin{equation}
    \label{eq:1}
    \sigma(\omega,\mu,\tau,T,B_0) = \hat{x}\hat{x}\sigma_{xx} + \hat{x}\hat{y}\sigma_{xy} + \hat{y}\hat{x}\sigma_{yx} + \hat{y}\hat{y}\sigma_{yy},\vspace*{10pt}
\end{equation}
where $\omega$, $\mu$, $\tau = 0.2 \, \text{ps}$ and $T = 300 \, \text{K}$ are angular frequency, chemical potential, temperature, and relaxation time, respectively. Moreover, dynamic  $(\sigma_{xx})$ and static $(\sigma_{xy})$ Hall conductivities are derived using the Kubo formalism as~\cite{KHODADADI2022443}:

\begin{align}
    \sigma_{xy}(\omega) &= \frac{i\hbar e^2}{L^2}  
    \sum_{\epsilon_\alpha<E_F} \sum_{\epsilon_\beta \geq E_F} 
    \frac{1}{\epsilon_\beta - \epsilon_\alpha} \times \nonumber \\
    &\quad \left( 
    \frac{j_x^{\alpha\beta} j_y^{\beta\alpha}}{\epsilon_\beta - \epsilon_\alpha - \hbar \omega} + 
    \frac{j_y^{\alpha\beta} j_x^{\beta\alpha}}{\epsilon_\beta - \epsilon_\alpha + \hbar \omega}
    \right)
    \label{eq:2}
\end{align}

\begin{equation}
    \sigma_{xx}(\omega) = 
    \frac{e^2 \hbar}{iS} 
    \sum_{\alpha,\beta} 
    \frac{ \left[ f(\epsilon_\alpha) - f(\epsilon_\beta) \right] 
           |⟨\alpha|\nu_x|\beta⟩|^2 }
         { (\epsilon_\alpha - \epsilon_\beta)
           (\epsilon_\alpha - \epsilon_\beta + \hbar\omega + i\delta) }.
    \label{eq:3} \vspace*{10pt}
\end{equation}
Here, $f(\epsilon)$, $\nu_x = \partial H / \partial p_x$, $S$, $L$, $E_F$, $\epsilon_{\alpha,\beta}$, $\delta$, $|\alpha,\beta\rangle$, $\hbar = 1.05 \times 10^{-34} \, \text{m}^2\text{kg/s}^2\text{K}$, $e = -1.6 \times 10^{-19} \, \text{C}$, and $H$ are the Fermi–Dirac distribution, velocity operator, area of the system, sample size, Fermi level (FE), system eigenenergy, positive infinitesimal, eigenstate, reduced Planck constant, electron charge, and Hamiltonian matrix, respectively~\cite{Moshiri_2021}. In the absence of a magnetostatic field ($B_0 = 0$), the off-diagonal conductivity terms ($\sigma_{xy}$ and $\sigma_{yx}$) in Eq.~(\ref{eq:2}) vanish, rendering the monolayer response isotropic. This isotropic conductivity response directly influences the optical behavior of graphene, particularly its effective permittivity and surface impedance, which are expressed as follows~\cite{KhodadadiNozhatNasari+2025}:
\begin{subequations}
\label{eq:4}
\begin{align}
    \epsilon_g &=1+i\frac{\sigma_{intra} +\sigma_{inter}}{\omega \epsilon_0 \Delta_g }, \label{eq:4a} \\
    Z_g &= \frac{1}{\sigma_{intra} +\sigma_{inter}}. \label{eq:4b}
\end{align}\vspace*{10pt}
\end{subequations}
where $\Delta_g$ is the effective thickness of graphene $(0.7~\text{nm})$; $\sigma_{\text{intra}}$ and $\sigma_{\text{inter}}$ represent the intraband and interband components of the conductivity, respectively~\cite{Khodadadi2023}:

\begin{subequations}
\label{eq:5}
\begin{align}
    \sigma_{\text{intra}} &= \frac{-ie^2 k_B T}{\pi \hbar^2 (\omega - i/\tau)} \left( \frac{\mu}{k_B T} + 2\ln\left(e^{-\mu / k_B T} + 1\right) \right), \label{eq:5a} \\
    \sigma_{\text{inter}} &\approx \frac{-ie^2}{4\pi \hbar^2} \ln \left( \frac{2|\mu| - (\omega - i/\tau)\hbar}{2|\mu| + (\omega - i/\tau)\hbar} \right). \label{eq:5b}
\end{align}\vspace*{10pt}
\end{subequations}

In the THz regime, $\hbar \omega \ll 2\mu$, making the interband, Eq.~(\ref{eq:5b}), contribution negligible due to Pauli blocking. Thus, graphene's conductivity simplifies to its intraband form:
\begin{equation}
    \label{eq:6}
    \sigma(\omega) \approx \sigma_{intra} = \frac{-ie^2\mu}{\pi \hbar^2(\omega - i/\tau)}.\vspace*{10pt}
\end{equation}

\begin{figure}
\centering
    \begin{subfigure}[b]{0.24\textwidth}
        \centering
        \includegraphics[width=\textwidth]{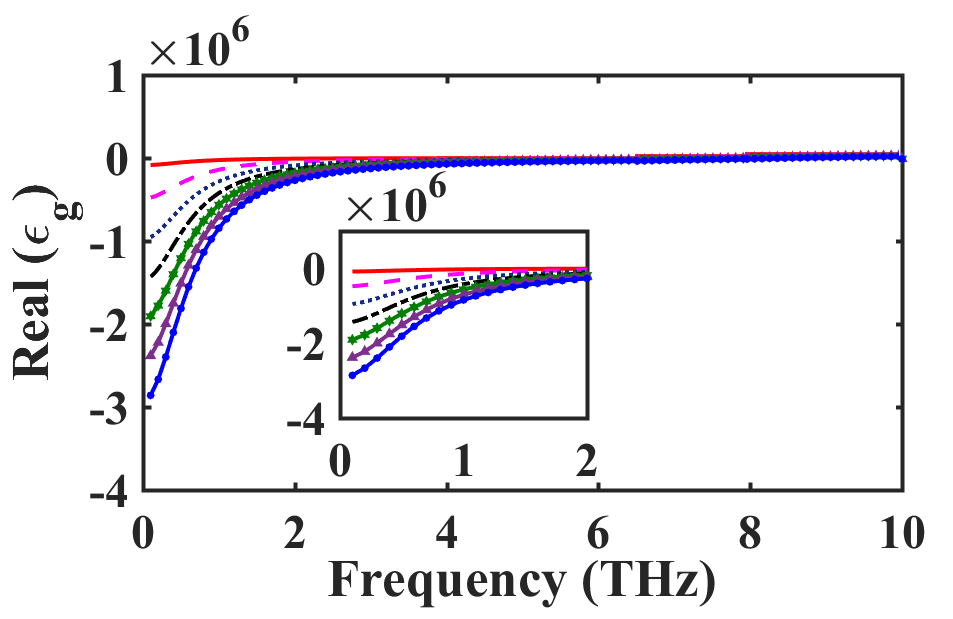}
        \caption{}
        \label{fig:fig2a}
    \end{subfigure}
    \begin{subfigure}[b]{0.24\textwidth}
        \centering
        \includegraphics[width=\textwidth]{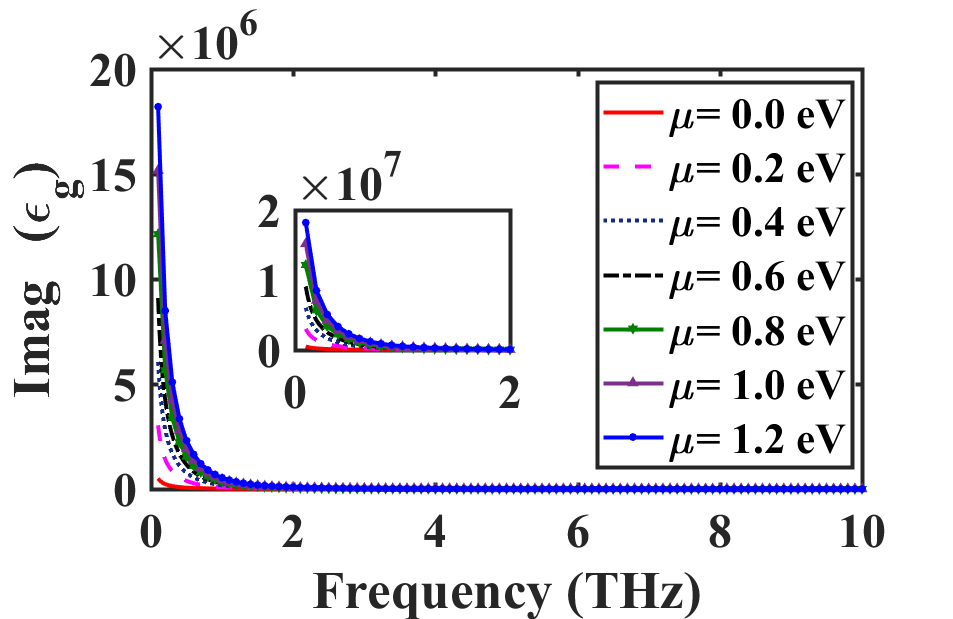}
        \caption{}
        \label{fig:fig2b}
    \end{subfigure}
    \caption{Permittivity spectra of graphene for varying chemical potentials. (a) Real and (b) imaginary parts of the permittivity spectra of graphene for different values of chemical potential. Insets highlight the response in the low-frequency range from $0$ to $2$~THz.}
    \label{fig:figure2}
\end{figure}

This tunable, low-loss response—enhanced by the transparency, insulation, and thermal robustness of the supporting $SiO_2$ substrate—creates an optimal environment for efficient THz plasmonic device operation.  To better understand this behavior, it is essential to analyze how graphene’s electromagnetic properties respond to changes in its chemical potential. Figure~\ref{fig:fig2a} and Figure~\ref{fig:fig2b} show the real and imaginary parts of graphene’s permittivity across various chemical potentials in the THz band, based on Eq.~(\ref{eq:5}). These spectra highlight how the dynamic tunability of graphene can be achieved by adjusting the chemical potential through chemical doping or electrostatic biasing. Such modulation not only alters the permittivity but also influences the conductivity (see Eq.~(\ref{eq:6})) and surface impedance (see Eq.~(\ref{eq:4})), thereby shifting the resonant characteristics and enabling controlled plasmonic behavior.

It is essential to mention that, achieving uniform control of the chemical potential $\mu$ across all pixels presents challenges, but several strategies can effectively minimize variation and enhance stability. Closed-loop control systems with integrated voltage sensors and feedback circuits can dynamically adjust gate voltages to compensate for drift and environmental changes. Incorporating low-noise voltage regulators and high-$k$ dielectrics further stabilizes electrostatic gating. Uniformity in the gate dielectric can be improved via atomic layer deposition (ALD) \cite{YuLiXiangLiCaoJiLiuXiaoYinGuoDai+2024+4019+4028}, reducing spatial $\mu$ variation. in Ref. \cite{doi:10.1021/nn504544h}, 
 Whitby shown that $\mu$ variations between 1.5–10 eV induce minimal shifts in the THz dielectric response. A 5–10\% variation in $\mu$ yields only marginal changes in the real and imaginary components of the effective permittivity (Figure~\ref{fig:figure2}), indicating a negligible impact on electromagnetic performance. This sensitivity lies well within the tolerance of standard gating methods. The integration of electronic control with plasmonic functionality thus enables scalable, adaptive THz platforms, where pixel-level tunability remains robust despite hardware imperfections.

\subsection{LSPR-Induced Virtual SPPs in Meta-Cells}

Beyond the tunable permittivity discussed previously, a fundamental mechanism underpinning graphene’s plasmonic response in the THz regime is the excitation of LSPRs. In the proposed architecture, LSPRs are excited by confined charge oscillations within subwavelength graphene meta-cell structures, which in our design have lateral dimensions of $14$~µm. The resonance condition is approximated by:
\begin{equation}
    \label{eq:eq7}
    L \approx \frac{m\pi}{\Re\{k_{\text{LSPRs}}\}},\vspace*{10pt}
\end{equation}
where $L$ represents the effective cavity length of the unit cell, $m$ is an integer mode number, and $k_{\text{LSPRs}}$ denotes the complex wavevector associated with the localized mode.

In this spectral range, graphene supports highly confined plasmonic modes due to its frequency-dependent surface conductivity $\sigma(\omega, \mu, \tau, T)$, enabling subwavelength confinement with $\lambda_{\text{LSPRs}} \ll \lambda_0$. The strong field localization facilitates enhanced light–matter interaction within each unit cell. Moreover, the tunability of the chemical potential $\mu$ via electrostatic gating allows dynamic control over the LSPR frequency and modal distribution. As a result, two-dimensional meta-pixel arrays are capable of supporting LSPRs under both transverse electric (TE) and transverse magnetic (TM) polarizations~\cite{SINGH2023115424}. When the chemical potential approaches zero ($\mu \approx 0$), interband electronic transitions dominate, leading to a negative imaginary component of conductivity. Under these conditions, TE-polarized LSPR modes can be supported, governed by the dispersion relation:
\begin{equation}
    \label{eq:eq8}
    k_{\text{LSPRs}}^{\text{(TE)}} = k_0 \sqrt{1 - \left( \frac{\sigma(\omega, \mu, \tau, T) n}{2} \right)^2},\vspace*{10pt}
\end{equation}

Conversely, for elevated chemical potentials ($\mu > 0$), intraband transitions become dominant, and the imaginary part of the conductivity becomes positive. In this regime, TM-polarized LSPR modes emerge, characterized by the nonlinear dispersion relation:
\begin{equation}
    \label{eq:eq9}
    \frac{1}{\sqrt{k_{\text{LSPRs}}^2 - \omega^2/c^2}} + \frac{\epsilon_r}{\sqrt{k_{\text{LSPRs}}^2 - \epsilon_r \omega^2/c^2}} = \frac{i \sigma(\omega, \mu, \tau, T)}{\omega \epsilon_0}.\vspace*{10pt}
\end{equation}

\begin{figure}
    \centering
    \includegraphics[trim={5cm 1cm 5cm 1cm},clip,width=0.24\textwidth]{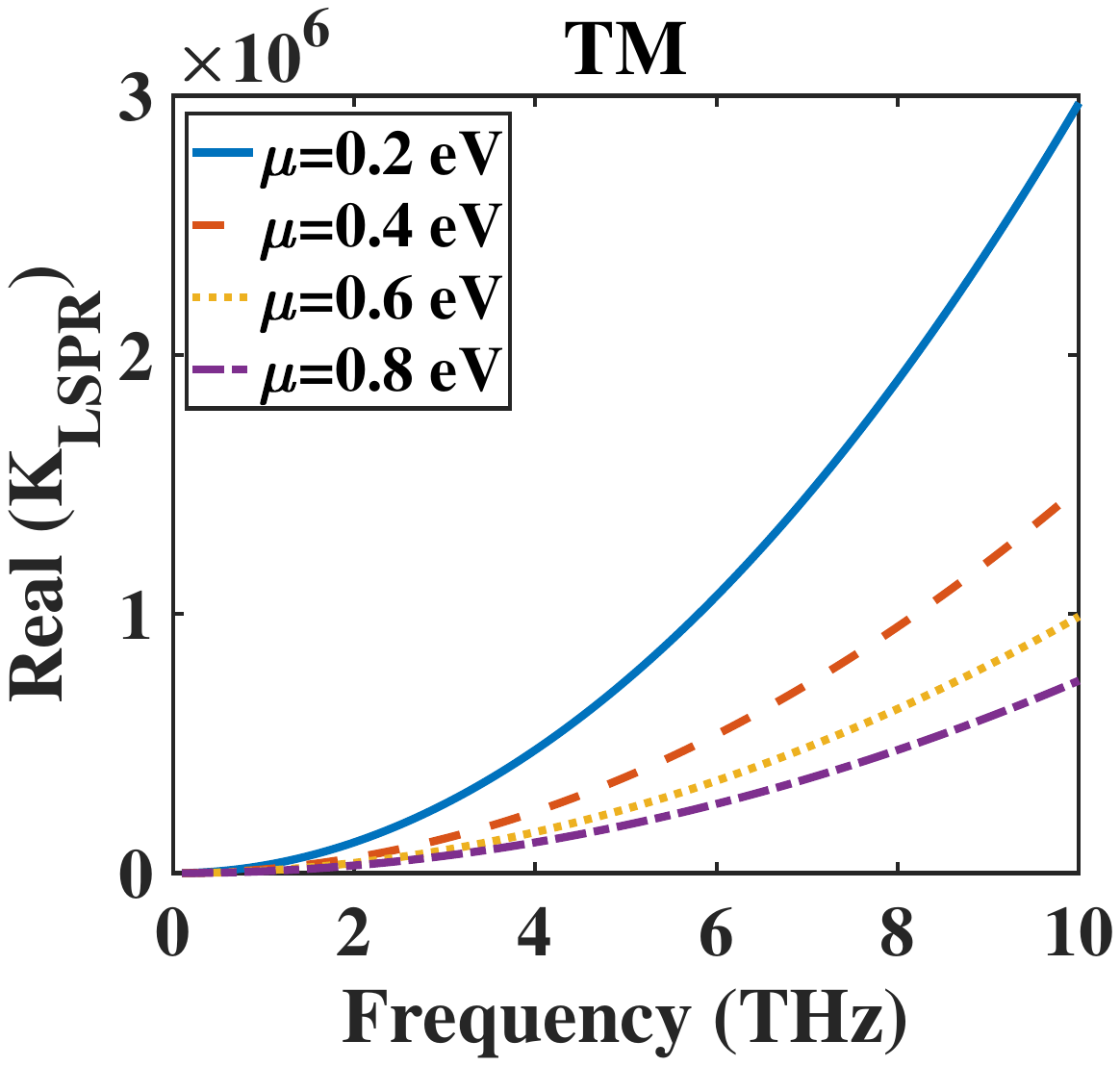}
    \includegraphics[trim={4.5cm 1cm 4.5cm 0cm},clip,width=0.24\textwidth]{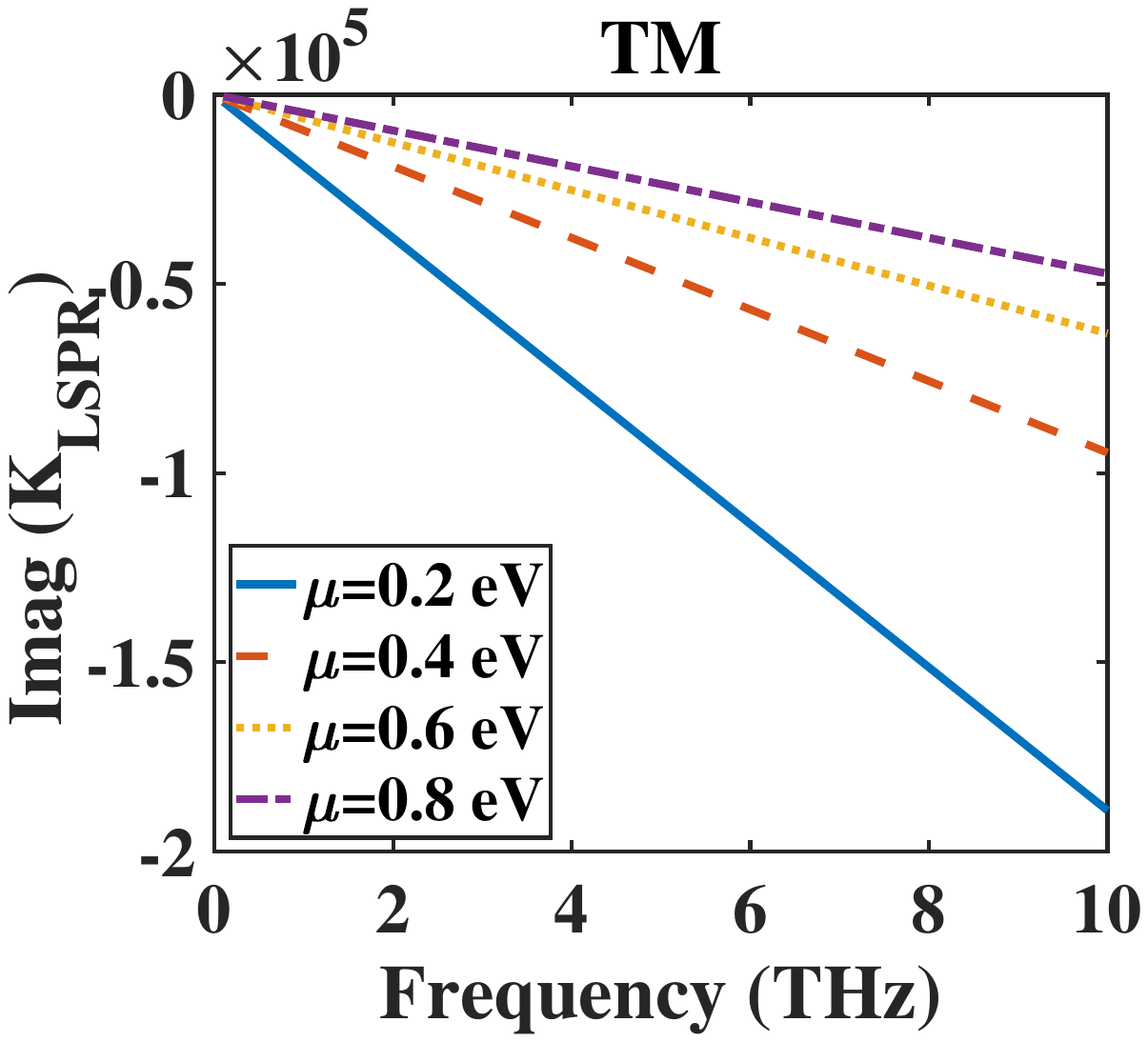}
    \includegraphics[trim={5.1cm 1cm 5cm 0cm},clip,width=0.24\textwidth]{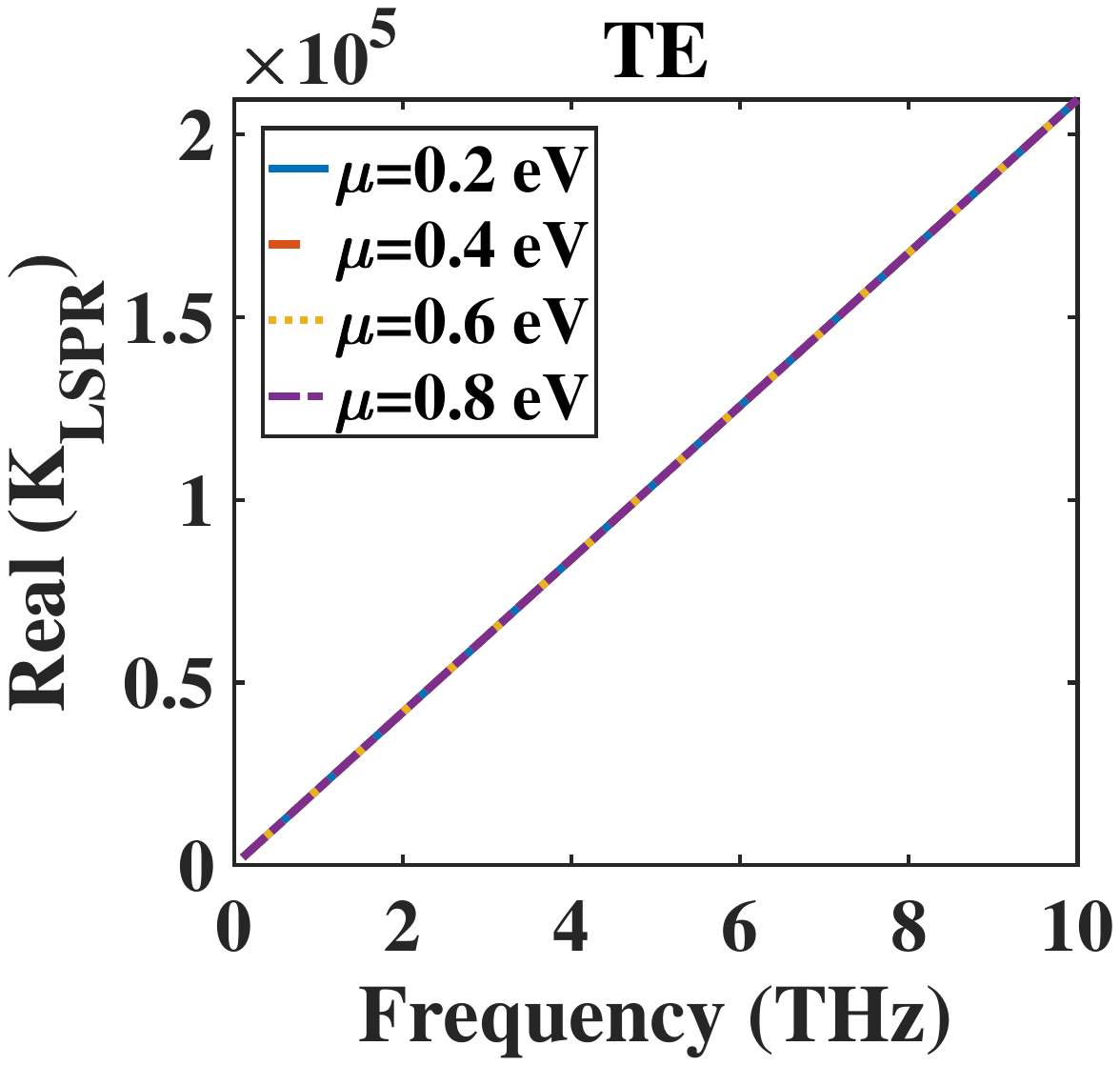}
    \includegraphics[trim={4.5cm 0.75cm 4.7cm 0cm},clip,width=0.24\textwidth]{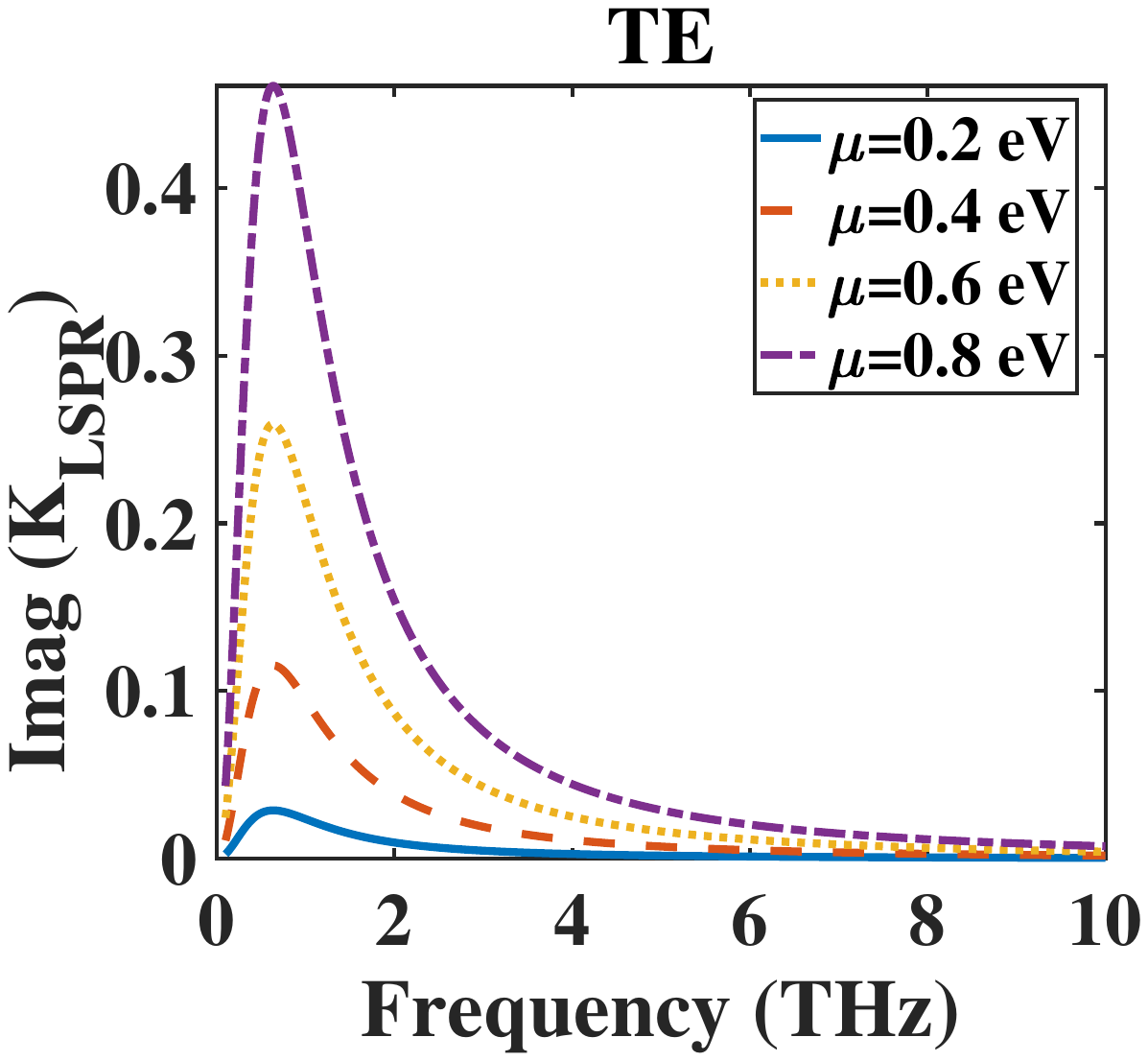}
    \caption{ Dispersion relation of LSPRs behaviour in graphene for both TM and TE modes as a function of the varied graphene chemical potential. Variation in the chemical potential of graphene influences the dispersion relation LSPRs for both TM  and TE modes. }
    \label{fig:fig5}
\end{figure}

Furthermore, the plasmon wavelength $\lambda_{\text{LSPRs}}$ for surface states in graphene can be analytically approximated as which help us to optimize the unit cell dimension:
\begin{equation}
    \label{eq:10}
    \lambda_{\text{LSPRs}} = \frac{\lambda_0 e^2}{\pi \epsilon_0 \hbar c} \cdot \frac{\mu}{\left(\epsilon_{r_1} + \epsilon_{r_2}\right)} \cdot \frac{1}{\hbar(\omega - i \tau^{-1})}. \vspace*{10pt}
\end{equation}
where $\epsilon_{r_1}$ and $\epsilon_{r_2}$ represent the dielectric constants of the upper and lower media surrounding the graphene sheet, respectively. Although initially derived for propagating SPPs in extended structures, Equations~(\ref{eq:eq8}) and~(\ref{eq:eq9}) are adapted here to describe localized plasmonic modes by treating finite graphene domains as effective plasmonic cavities. To resolve the transcendental nature of Eq.~(\ref{eq:eq9}), a genetic algorithm was employed, offering robust convergence across complex, multidimensional parameter spaces. The resulting values of $k_{\text{LSPRs}}$ validate the resonance condition in Eq.~(\ref{eq:eq7}), demonstrating that graphene meta-cells of $14$~µm can effectively support tunable, compact LSPR modes throughout the THz frequency range. The influence of chemical potential variation on the excited LSPRs is depicted in Figure \ref{fig:fig5}, confirming the correlation between chemical potential and SPPs behaviour of graphene for both TM and TE modes. 
Moreover, the linear dispersion relation near the Dirac points in the graphene band structure facilitates the coherent oscillation of its surface electrons upon interaction with THz EM waves, resulting in LSPRs excitation \cite{SINGH2023115424}. 

\begin{figure*}
\centering
    \begin{subfigure}[b]{\textwidth}
        \centering
        \includegraphics[trim={0cm 10.1cm 0cm 0cm},clip,width=0.8\textwidth]{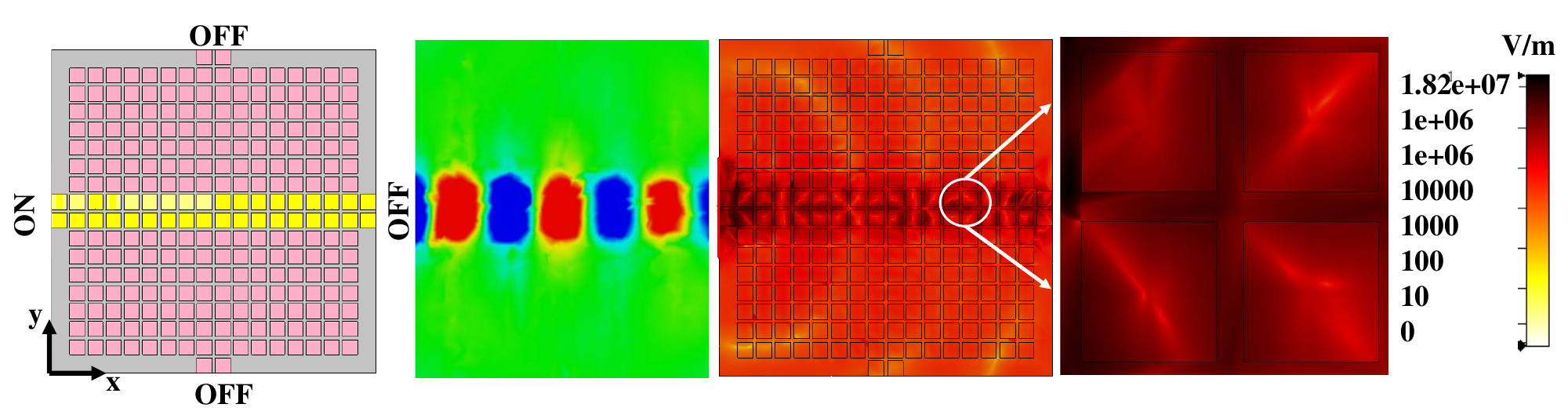}
        \caption{}
        \label{fig:fig6a}
    \end{subfigure}
    \begin{subfigure}[b]{\textwidth}
        \centering
        \includegraphics[trim={0cm 10.8cm 0cm 1cm},clip,width=0.8\textwidth]{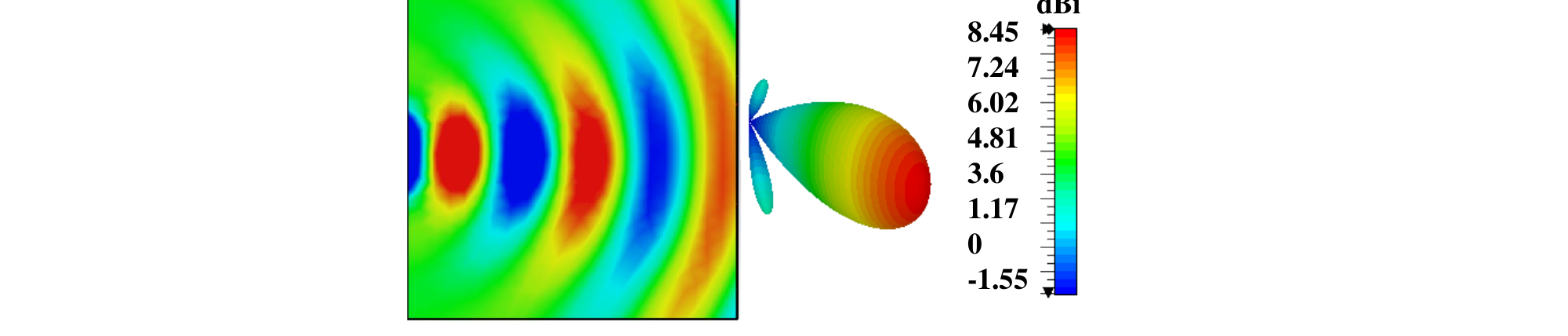}
        \caption{}
        \label{fig:fig6b}
    \end{subfigure}
    \caption{Excitation of virtual SPPs via near-field coupling to LSPRs. (a) Closed waveguide configuration showing the $E_z$ field distribution on the $x$-$y$ plane, highlighting LSPR and virtual SPP excitation across the structure and in zoomed views. (b) Open waveguide configuration, where an open add-space boundary enables virtual SPPs to radiate into free space, forming a directional radiation pattern.}
    \label{fig:fig6}
\end{figure*}

To better understand the excitation of LSPRs and the emergence of virtual SPPs in the THz regime, we consider a waveguide-fed metasurface composed of graphene meta-pixels. Near the excitation ports, localized surface plasmon resonances are first triggered within individual meta-cells. These strongly confined fields do not remain isolated; rather, they interact with neighboring meta-pixels through near-field coupling. This inter-cell coupling enables the energy of the initial LSPRs excitation to propagate across adjacent unit cells, creating a cascading chain of resonant interactions. This dynamic behavior mimics the characteristics of propagating SPPs, despite being formed through discrete localized interactions — hence the term “virtual SPPs.” In essence, the LSPRs excitations hop from one meta-pixel to the next, giving rise to a wave-like propagation of energy across the metasurface, without the need for continuous material interfaces typical of classical SPPs. This chain effect transforms the local LSPRs into a collective mode that extends across multiple cells. Figure~\ref{fig:fig6}(a) visually captures this process: the initial LSPR excitation couples to adjacent cells, forming a coherent virtual SPPs channel. As shown in Figure~\ref{fig:fig6}(b), this virtual mode not only propagates but also radiates into free space, acting as an effective meta-antenna. The ability of virtual SPPs to support both guided and radiative behavior opens up new opportunities for designing compact, efficient THz plasmonic emitters and modulators, where the structure itself serves as both waveguide and antenna.

\section{Binary Field-Driven Meta-Routing Method}

To enable adaptive on-chip THz wireless communication, we introduce a field-driven meta-routing method that leverages binary control over graphene meta-pixels to dynamically shape energy pathways. In their active state, pixels support strong plasmonic resonances and enable energy transfer; in the passive state, they suppress plasmon excitation and behave like reflective or inactive nodes, redirecting or blocking the field. By selectively activating these pixels, near-field coupled chains of localized surface plasmons are formed—creating virtual SPP channels that guide energy along programmable routes. These reconfigurable antenna pathways enable directional emission, dynamic beam steering, polarization control, and efficient programmable multi-beam routing in densely integrated chip-scale networks.
\begin{figure}
    \centering
    \includegraphics[width=0.5\textwidth]{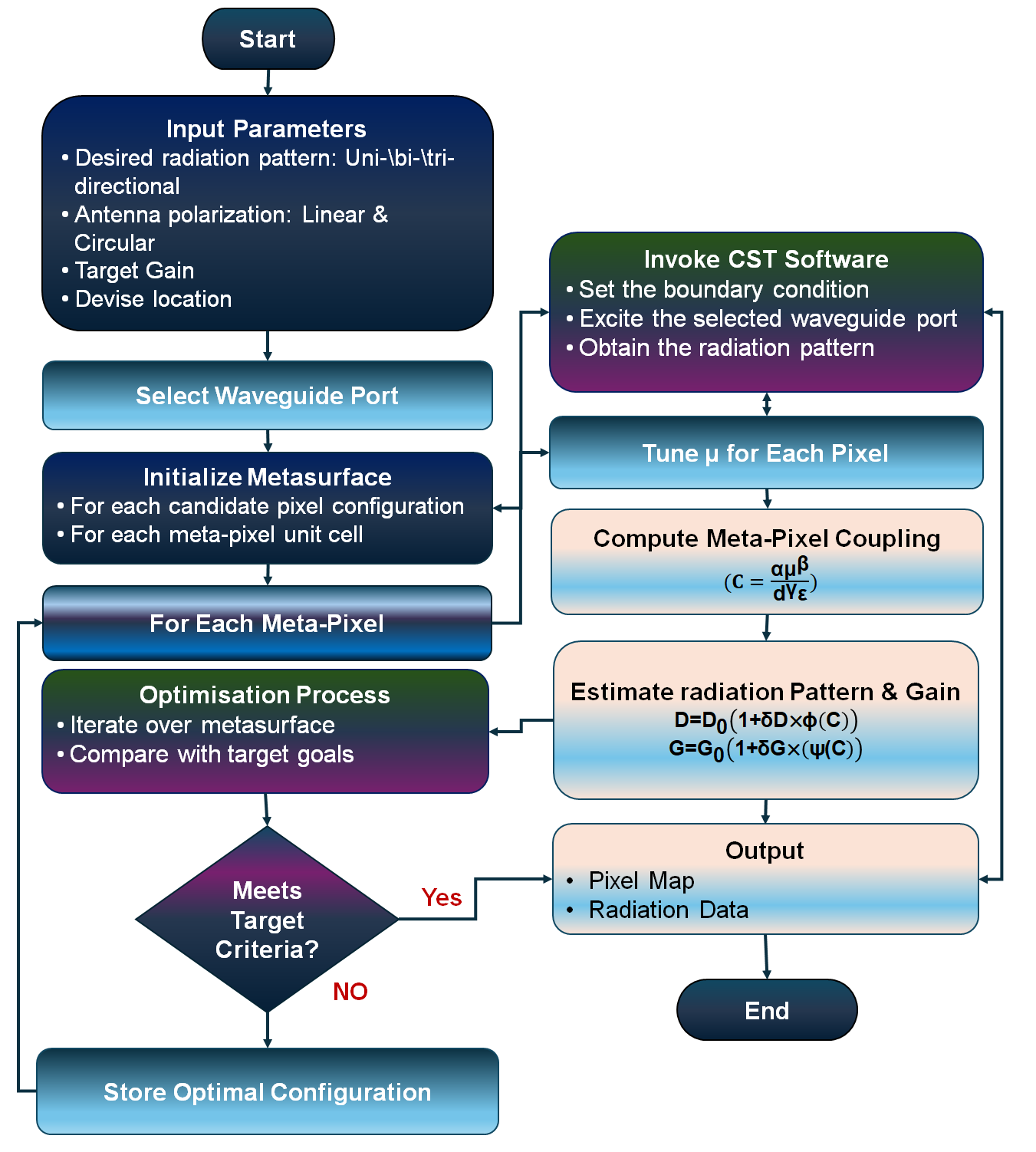}
    \caption{Flowchart of binary field-driven meta-routing method to dynamically shape energy pathways of LSPRs to create virtual SPP channels. This is achieved through impedance modulation of meta-pixel by altering its chemical potential.}
    \label{fig:fig7}
\end{figure}
To model the impact of pixel activation on system performance, the algorithm—illustrated in the flowchart of Figure~\ref{fig:fig7}—analyzes the coupling between LSPRs in graphene meta-pixels and their adjacent unit cells. By tuning the chemical potential \( \mu \), each pixel alternates between metallic and dielectric behavior, facilitating the controlled formation of these virtual SPP pathways. The coupling strength is estimated as:
\begin{equation}
C = \frac{\alpha \cdot \mu^\beta}{d^\gamma \cdot \epsilon},\vspace*{10pt}\label{eq:eq11}
\end{equation}
where \( d \) is the inter-cell spacing, \( \epsilon \) is the local permittivity, and \( \alpha, \beta, \gamma \) are empirical constants. This coupling influences both the antenna’s radiation directivity and gain, which are modeled as:
\begin{align}
D &= D_0 \cdot \left(1 + \delta_D \cdot \phi(C)\right), \\
G &= G_0 \cdot \left(1 + \delta_G \cdot \psi(C)\right).\vspace*{10pt}
\end{align}
where \( D_0 \) and \( G_0 \) are baseline values, and \( \phi(C), \psi(C) \) describe how coupling modulates system behavior—derived from simulations or empirical fitting. To supress side lobes, an optimization routine is applied to the spatial distribution of 
$\mu$, leveraging its effect on the coupling strength $C$. By tuning $\mu$ across the pixel array, 
destructive interference in off-axis directions is enhanced, while constructive interference 
is preserved along the main beam. Gradient-based or evolutionary algorithms are employed to 
iteratively adjust the local $\mu$ distribution, minimizing sidelobe levels while adhering 
to physical and fabrication constraints.

\section{Coupled-Mode Theory of Field-Driven LSPR Meta-Networks}

As illustrated in Figure~\ref{fig:fig1}, each meta-pixel unit cell within the metasurface supports a LSPR mode, represented by a complex amplitude $a_{mn}(t)$ at lattice site $(m,n)$. The resonant behavior of each unit cell is actively tunable through the local graphene chemical potential $\mu_{mn}$, which directly modulates the surface impedance $Z_{mn}$. The resonance frequency $\omega_{mn}$ of each LSPR is assumed to depend linearly on the inverse surface impedance, such that:

\begin{equation} \omega_{mn} = \omega_{\text{ref}} + \beta \left( \frac{1}{Z_{mn}} - \frac{1}{Z_{\text{ref}}} \right), \label{eq:eq14}\end{equation}
where $\omega_{\text{ref}}$ is the reference frequency corresponding to a known impedance $Z_{\text{ref}}$, and $\beta$ is a tuning sensitivity parameter determined by the structural and material properties of the meta-pixel. To estimate $\beta$, we model each graphene patch as a microstrip-like resonator with effective cavity length $L_{\text{eff}}$ and effective dielectric constant $\epsilon_{\text{g}}$. The unperturbed resonance frequency is approximated by:

\begin{equation} \omega_0 = \frac{c}{L_{\text{eff}} \sqrt{\epsilon_{\text{g}}}}, \end{equation}
where $c$ denotes the speed of light in vacuum. When the graphene’s surface impedance $Z_g$ is considered as a reactive boundary condition, it perturbs the effective cavity length and modifies the resonant frequency. Using first-order perturbation theory, the resonance frequency as a function of the chemical potential is:

\begin{equation} \omega_0(\mu_{mn}) \approx \omega_{\text{ref}} + \frac{\partial \omega_0}{\partial Z_g} \cdot \frac{\partial Z_g}{\partial \mu_{mn}} \cdot \Delta \mu_{mn}. \end{equation}

Accordingly, the sensitivity coefficient $\beta$ can be expressed as:

\begin{equation} \beta = \frac{\partial \omega_0}{\partial (1/Z_g)} \approx -\omega_{\text{ref}} \cdot \frac{Z_g^2}{L_{\text{eff}} Z_0}, \label{eq:eq17} \end{equation}
where $Z_0$ is the characteristic impedance of the surrounding medium or guiding structure. With the site-dependent resonance frequencies established, the spatiotemporal evolution of modal amplitudes across the meta-network is governed in the steady state by frequency-domain coupled-mode theory:

\begin{equation} \left( \omega - \omega_{mn} + j\gamma \right) a_{mn} - \kappa \sum_{(i,j) \in \text{NN}(m,n)} a_{ij} = s_{mn},\label{eq:eq18} \end{equation}
where $\gamma$ represents the intrinsic damping factor, $\kappa$ is the nearest-neighbor coupling coefficient, and $s_{mn}$ is the external source term applied to selected unit cells. The summation spans the set of nearest neighbors $\text{NN}(m,n)$ of each cell. Assuming a waveguide port excites the structure at two adjacent meta-pixels, specifically at $(\frac{m}{2}-1,\frac{n}{2}-1)$ and $(\frac{m}{2},\frac{n}{2}-1)$, the source term is defined as:

\begin{equation}
s_{mn} = 
\begin{cases}
S_0, & \text{if } (m,n) \in 
\left\{ 
\left(\frac{m}{2}-1,\frac{n}{2}-1\right), 
\left(\frac{m}{2},\frac{n}{2}-1\right) 
\right\} \\
0, & \text{otherwise}
\end{cases}
\end{equation}

To solve the coupled-mode equation in Eq.~(\ref{eq:eq18}), three key parameters must be determined: the local resonant frequency \( \omega_{m,n} \), the intrinsic damping factor \( \gamma \), and the inter-site coupling coefficient \( \kappa \). The value of \( \omega_{m,n} \) is obtained from Eq.~(\ref{eq:eq14}), which links the resonance to the local surface impedance \( Z_{m,n} \), itself modulated by the graphene chemical potential \( \mu_{m,n} \) via Eq.~(\ref{eq:4b}). 

The sensitivity coefficient \( \beta \) in Eq.~(\ref{eq:eq14}) is calculated from the microstrip resonator model using Eq.~(\ref{eq:eq17}), where the effective dielectric constant \( \epsilon_g \) is extracted from the permittivity expression in Eq.~(\ref{eq:4a}). The damping factor \( \gamma \) is determined from the imaginary part of the complex resonance frequency, which is influenced by both intrinsic graphene losses (via the scattering rate \( 1/\tau \)) and radiation leakage at the boundary of the meta-array. We estimate \( \gamma \) by fitting the theoretical \( S_{11} \) to full-wave results, ensuring accurate impedance bandwidth matching. Finally, the coupling coefficient \( \kappa \) is estimated using Eq.~(\ref{eq:eq11}), based on inter-pixel spacing, local permittivity, and empirical tuning. These values are either computed using physical constants or calibrated via comparison with full-wave numerical results, ensuring the theoretical model faithfully captures the spatial dynamics of the metasurface.

To facilitate theoretical analysis and capture the collective interactions across the entire array, the coupled-mode equations can be reformulated in compact matrix form:
\begin{equation}
\mathbf{M} \cdot \mathbf{a} = \mathbf{s},
\end{equation}
where $\mathbf{a} \in \mathbb{C}^{256}$ is the vector of complex modal amplitudes, $\mathbf{s}$ is the source vector, and $\mathbf{M} \in \mathbb{C}^{256 \times 256}$ is the system matrix with entries:
\begin{equation}
\begin{aligned}
M_{(m,n),(m,n)} &= \omega - \omega_{mn} + j\gamma, \\
M_{(m,n),(i,j)} &= -\kappa, \quad \text{if } (i,j) \in \text{NN}(m,n).
\end{aligned}
\end{equation}

To quantify the reflection response at the excitation port, the input reflection coefficient $S_{11}$ can be derived from the ratio of reflected to incident power at the excited meta-pixels. Using the local amplitudes $a_{mn}$ at the excitation sites, the complex surface current at each meta-pixel is given by $J_{mn} = a_{mn} / Z_{mn}$, leading to:

\begin{equation}
S_{11} = \sqrt{\frac{\sum_{(m,n) \in \text{port}} \left| \frac{a_{mn}}{Z_{mn}} \right|^2}{|S_0|^2 / Z_{\text{port}}}},
\end{equation}
where $Z_{\text{port}}$ is the characteristic impedance of the feeding waveguide. Beyond reflection characteristics, the same modal amplitudes $a_{mn}$ can be used to compute the surface currents responsible for far-field radiation, which in turn allows evaluation of the total electric field radiated in the direction $(\theta, \phi)$ using the array factor expression:

\begin{equation} E(\theta, \phi) \propto \sum_{m=0}^{15} \sum_{n=0}^{15} J_{mn} \cdot e^{j k \left( m d_x \sin\theta \cos\phi + n d_y \sin\theta \sin\phi \right)},\label{eq:eq23} \end{equation}
where $k$ is the free-space wavenumber, and $d_x$, $d_y$ are the lattice constants in the $x$ and $y$ directions, respectively. Based on the radiated electric field distribution, the directional gain pattern can be derived by normalizing the radiation intensity:

\begin{equation} G(\theta, \phi) = 10 \log_{10} \left( \frac{|E(\theta, \phi)|^2}{\max_{\theta,\phi} |E(\theta, \phi)|^2} \right) + G_{\text{max}}, \label{eq:eq24}\end{equation}
The term \( E(\theta, \phi) \) in Eq.~(\ref{eq:eq23}) reflects the coherent superposition of radiated fields from excited meta-pixels. In practice, only a subset of these contribute significantly to radiation --- primarily those near the boundary of the virtual SPP path, where near-field to far-field coupling occurs due to impedance discontinuity and reduced confinement.
Also, the maximum achievable gain $G_{\text{max}}$ is approximated by:

\begin{equation} G_{\text{max}} = 10 \log_{10} \left( \frac{4\pi}{\Omega_{\text{beam}}} \right), \end{equation}
and $\Omega_{\text{beam}}$ is the beam's solid angle. Finally, to further characterize the polarization properties of the radiated field, the axial ratio (AR) can be computed by decomposing the far-field electric field into orthogonal components, such as $E_\theta$ and $E_\phi$, at a given observation angle. The AR is defined as:

\begin{equation}
\text{AR}(\theta, \phi) = \frac{ \left| |E_\theta| + j |E_\phi| \right| }{ \left| |E_\theta| - j |E_\phi| \right| },
\end{equation}
An AR close to unity (0 dB), or typically less than 3 dB,indicates circular polarization (CP), while larger values correspond to elliptical or linear polarization. This analysis facilitates the identification of the proposed controllable on-chip antenna polarization states enabled by the field-driven metasurface. In the following section, we compare the numerical results with theoretical predictions to demonstrate the superior performance of the proposed field-driven meta-pixel architecture for on-chip communication.

\section{Results and Discussions}
Although SPPs are intrinsically non-radiative, in the proposed architecture, radiation is enabled through a combination of near-field LSPR coupling and controlled termination. Specifically, the proposed \textit{ Binary Field-Driven Meta-Routing Method} selectively activates a subset of graphene meta-pixels, shaping localized plasmonic excitations into a coherent virtual SPP channel. As these channels propagate and reach discontinuities — particularly at the boundary of the activated region — the lack of confinement and impedance mismatch leads to radiation into free space. This mechanism resembles that of leaky-wave antennas and synthetic aperture emitters, where constructive interference among phase-coherent LSPRs results in directional beam formation. Notably, radiation primarily originates from the subset of patches near the aperture edge rather than uniformly from all pixels with $\mu>=0$, enabling precise beam steering and suppression of side lobes. Therefore, in this section, the effectiveness of the proposed \textit{Binary Field-Driven Meta-Routing Method} in enabling dynamic control over electromagnetic wave propagation within a graphene-based metasurface network is demonstrated. By locally tuning the chemical potential of each unit cell, distinct routing topologies—supporting unidirectional, bidirectional, and multidirectional (tri-directional) radiation modes—are configured. Programmable control over polarization states, including both linear and circular polarizations, is also achieved. These configurations are considered critical for the realization of reconfigurable on-chip wireless communication systems, where dynamic beam steering and polarization multiplexing contribute to enhanced data throughput, spatial reuse, and interference mitigation \cite{YAO2021400, WANG201842, DEHDAST2021625}. Theoretical predictions derived from the \textit{Coupled-Mode Theory of Field-Driven LSPR Meta-Networks} are validated through full-wave numerical simulations. The strong agreement between the results serves as a proof of concept for the feasibility and accuracy of the proposed meta-routing framework, highlighting the potential of chemically tunable graphene metasurfaces as a scalable and compact platform for THz integrated wireless networks.

\subsection{Unidirectional Recursive Beam Steering via Meta-Routing}

\begin{figure}
\centering
        \begin{subfigure}[b]{0.5\textwidth}
         \centering
         \includegraphics[width=\textwidth]{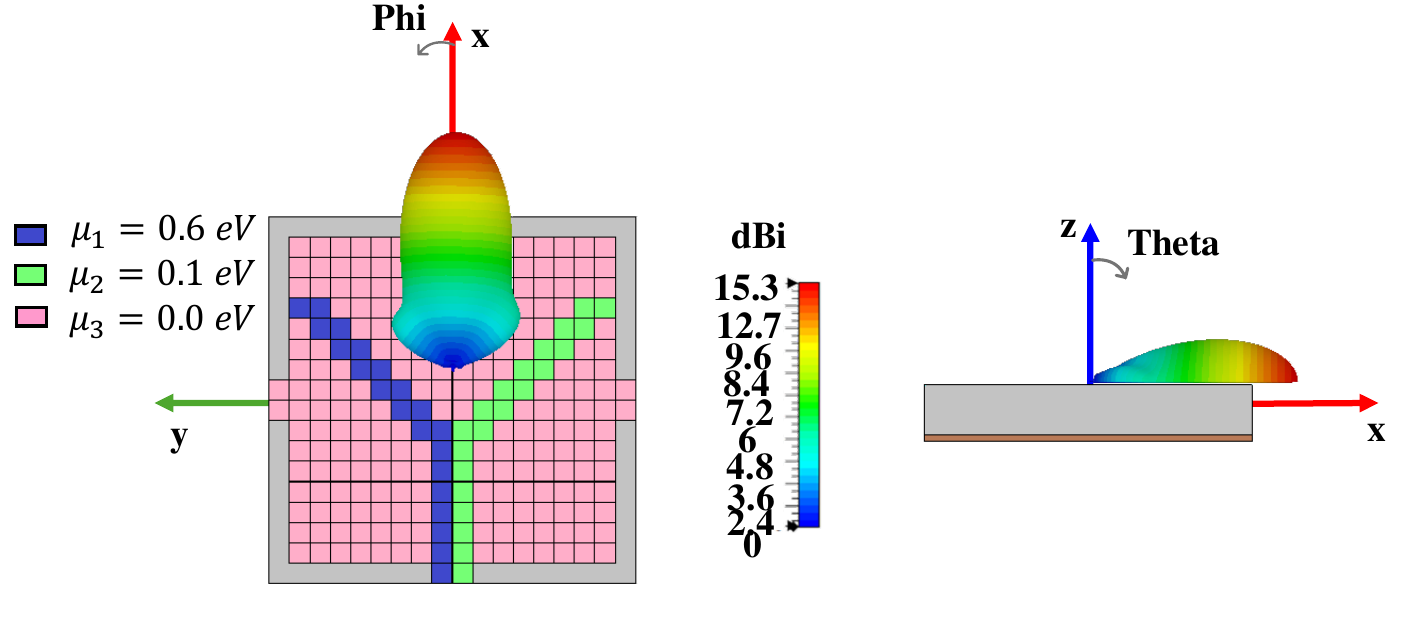}
         \caption{}
         \label{fig:fig8a}
     \end{subfigure}
     \begin{subfigure}[b]{0.24\textwidth}
         \centering
         \includegraphics[trim={3cm 9cm 4cm 9.5cm},clip,width=\textwidth]{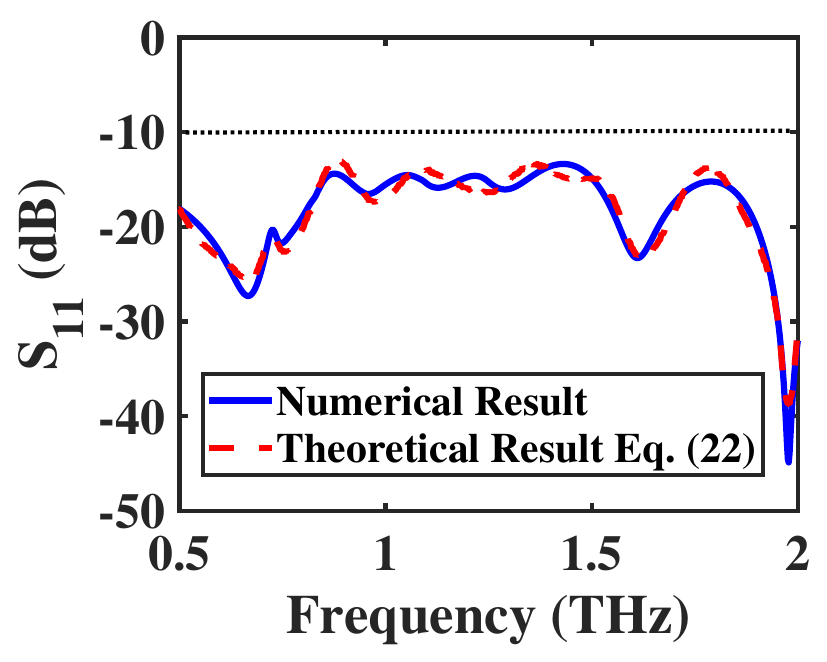}
         \caption{}
         \label{fig:fig8b}
     \end{subfigure}
    \begin{subfigure}[b]{0.24\textwidth}
         \centering
         \includegraphics[trim={3cm 9cm 3.7cm 9.5cm},clip,width=\textwidth]{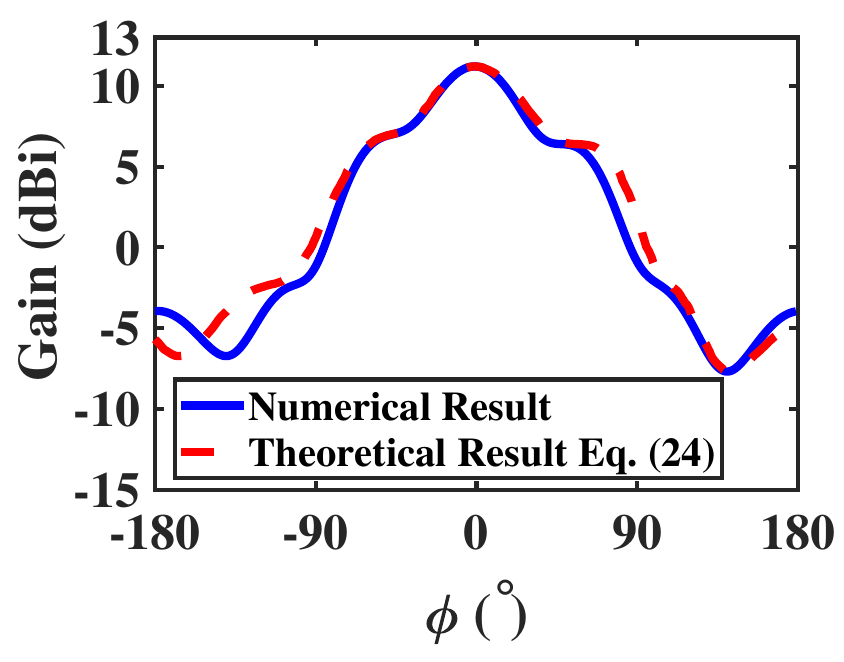}
         \caption{}
         \label{fig:fig8c}
     \end{subfigure}
    \caption{
Theoretical and full-wave numerical performance of the proposed \textit{Y-MetaRouter}. 
(a) Top and side schematic views of the Y-MetaRouter, along with the simulated 2D far-field radiation pattern at $1.8$~THz. 
(b) Theoretical and simulated reflection coefficient ($S_{11}$) spectra, demonstrating impedance matching and resonant behavior. 
(c) One-dimensional (1D) gain radiation pattern at $\theta = 90^\circ$ and frequency of $1.8$~THz.
}
    \label{fig:fig8}
\end{figure}

\textbf{Y-MetaRouter Configuration:} Figure~\ref{fig:fig8} illustrates the theoretical and numerical performance of the proposed Y-MetaRouter, designed to achieve unidirectional beam steering through a spatially programmed graphene metasurface. By employing the \textit{Binary Field-Driven Meta-Routing Method}, the antenna is configured to generate a horizontally directed surface radiation pattern by selectively biasing regions of the meta-pixel networks with distinct chemical potentials. Specifically, the left arm of the Y-MetaRouter is doped with $\mu_1 = 0.6$~eV, enabling strong excitation of LSPRs, while the right arm is lightly doped with $\mu_2 = 0.1$~eV to balance the field distribution. The remaining meta-pixel unit cells are unbiased ($\mu_3 = 0.0$~eV), functioning as passive dielectric regions that suppress unwanted excitation and aid in shaping the beam. The resulting current distribution yields a stable, unidirectional main lobe at $0^\circ$, as confirmed by both theoretical predictions and full-wave simulations. The design achieves excellent impedance matching, with a reflection coefficient ($S_{11}$) below $-10$~dB between $0.5$ to $2$~THz. The gain, directivity, and total radiation efficiency are recorded as $15.1$~dBi, $15.3$~dBi, and $-0.2$~dB, respectively, while maintaining a main beamwidth of $65.4^\circ$ and side lobe suppression of $-11.1$~dB at the frequency of $1.8$~THz.

\begin{figure}
\centering
         \begin{subfigure}[b]{0.28\textwidth}
         \centering
         \includegraphics[trim={4.5cm 9cm 4.5cm 9cm},clip,width=\textwidth]{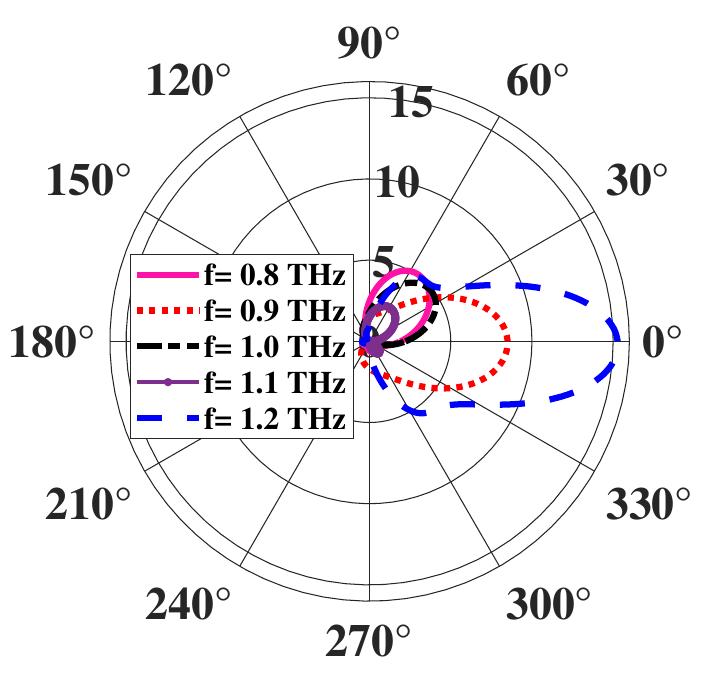}
         \caption{}
         \label{fig:fig9a}
     \end{subfigure}
              \begin{subfigure}[b]{0.46\textwidth}
         \centering
         \includegraphics[trim={1cm 1.8cm 3.1cm 0cm},clip,width=\textwidth]{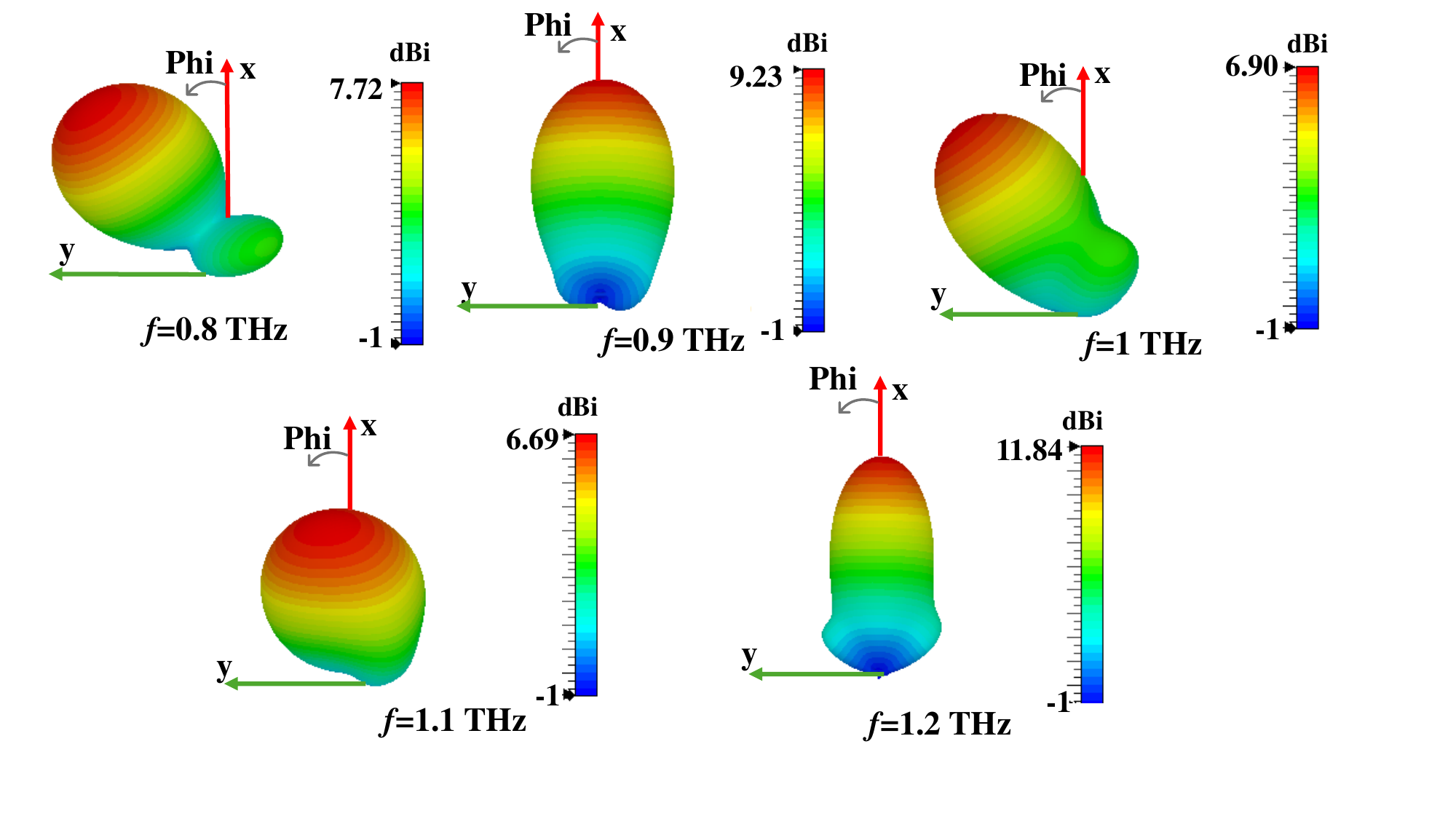}
         \caption{}
         \label{fig:fig9b}
     \end{subfigure}
  \caption{
Recursive beam steering performance of the proposed \textit{Y-MetaRouter} as the Meta-Gateway link. 
(a) Theoretical polar plots and (b) full-wave numerical far-field radiation patterns corresponding to the configuration shown in Figure~\ref{fig:fig8}, evaluated at different frequencies.} \label{fig:fig9}
\end{figure}

\textbf{Meta-Gateway Link Performance}: Figure~\ref{fig:fig9} illustrates the frequency-dependent beam steering behavior of the Y-MetaRouter, demonstrating its dynamic reconfigurability across the THz spectrum. Figure~\ref{fig:fig9}(a) presents theoretical polar radiation patterns derived from Eq.~(\ref{eq:eq24}), while Figure~\ref{fig:fig9}(b) shows the corresponding full-wave numerical far-field results. As the operating frequency increases from $0.8$ to $1.2$~THz, the main lobe exhibits a systematic angular shift—characteristic of frequency scanning antennas—highlighting the effectiveness of the recursive beam steering mechanism. This behavior arises from frequency-dependent modulation of surface impedance and effective permittivity in the graphene meta-pixels, which alters the local resonance conditions and wavevector of the excited LSPRs within the meta-network. At lower frequencies (e.g., $0.8$~THz), the beam is steered toward approximately $55^\circ$, whereas at higher frequencies (e.g., $1.2$~THz), it approaches the broadside direction. This frequency-scanning beam steering offers advantages for dynamic beam targeting and spectral multiplexing, with beam angle variability across the operating bandwidth enhancing adaptability in diverse communication scenarios.

\textbf{MetaSwitcher Configuration}: 
While the results presented so far demonstrate beam steering as a function of excitation frequency, it is important to distinguish this from the dynamic reconfigurability enabled by active biasing of the graphene meta-pixels. Frequency-driven beam steering, though effective, inherently suffers from direction–frequency coupling, which can degrade performance for wideband signals due to spatial dispersion or beam squinting. To fully realize the promise of programmable on-chip communication, it is desirable to achieve beam reconfiguration at a fixed frequency by dynamically altering the chemical potential distribution.

\begin{figure}
\centering
         \begin{subfigure}[b]{0.21\textwidth}
         \centering
         \includegraphics[trim={10cm 2cm 10cm 2cm},clip,width=\textwidth]{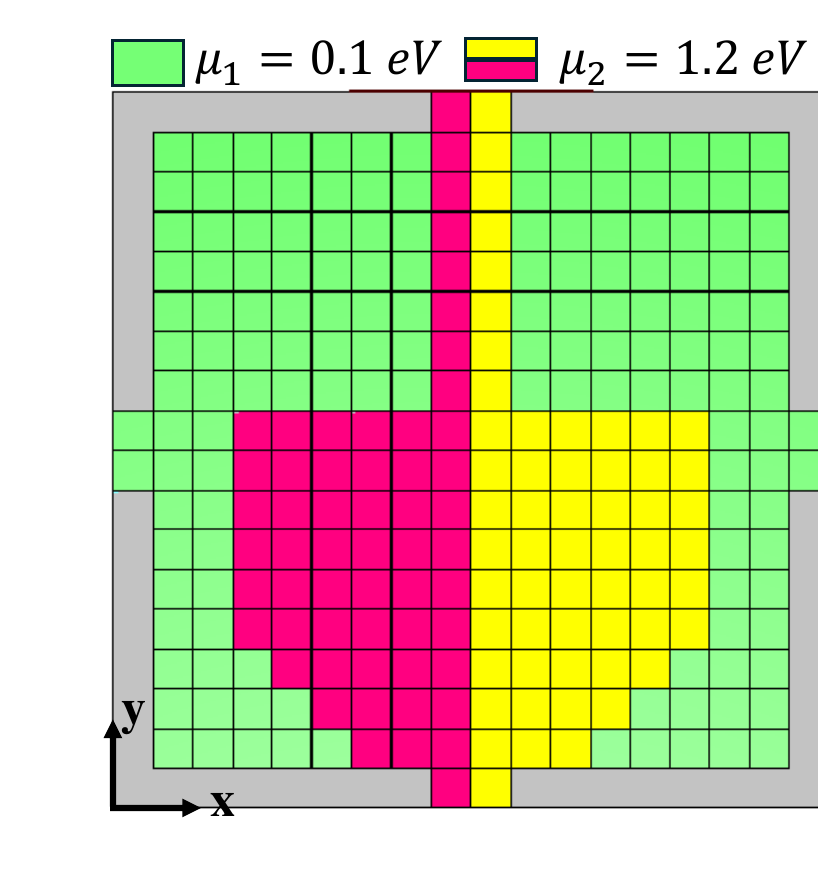}
         \caption{}
         \label{fig:fig10a}
     \end{subfigure}
              \begin{subfigure}[b]{0.27\textwidth}
         \centering
         \includegraphics[trim={3.2cm 9.5cm 3.2cm 9cm},clip,width=\textwidth]{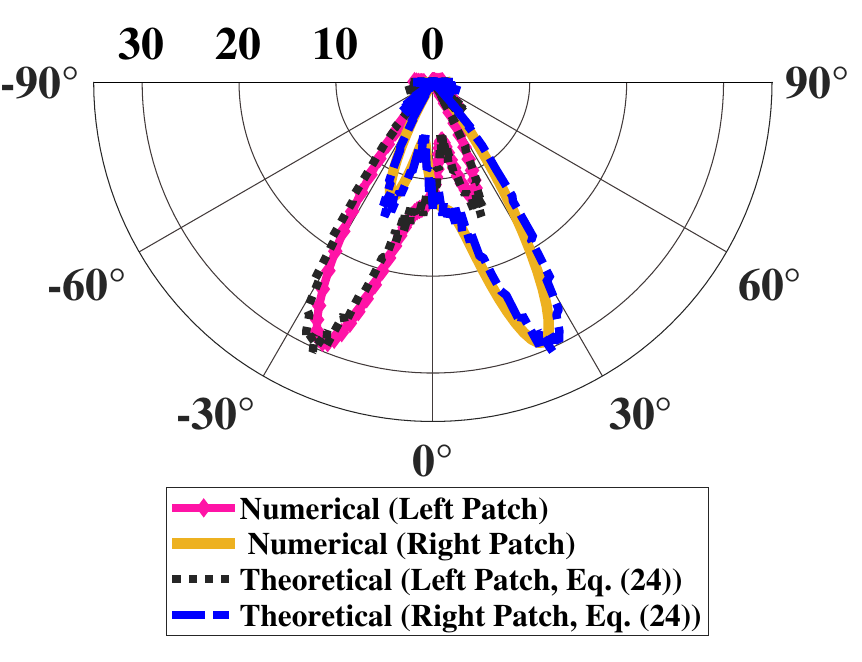}
         \caption{}
         \label{fig:fig10b}
     \end{subfigure}
\caption{
Radiation performance of the proposed \textit{MetaSwitcher} antenna to achieve beam reconfiguration at a fixed frequency. 
(a) Schematic top view of the graphene meta-network. 
(b) Theoretical and numerical polar plots of the far-field radiation pattern at 2.7~THz under different switching scenarios, achieved via the \textit{Binary Field-Driven Meta-Routing Method}. 
This figure highlights the MetaSwitcher’s ability to dynamically steer radiation by activating or deactivating individual patches, enabling highly directional and reconfigurable THz beam control.
}
    \label{fig:fig10}
\end{figure}

Figure~\ref{fig:fig10} presents the theoretical and numerical performance of the proposed \textit{MetaSwitcher} antenna, a novel structure designed for dynamically steering THz radiation through selective activation and deactivation of graphene meta-pixels. As shown in Figure~\ref{fig:fig10}(a), the configuration consists of a semi-patch metasurface, where chemical potential modulation defines the active and inactive regions, enabling binary pixel control. To achieve unidirectional beam steering, the Binary Field-Driven Meta-Routing Method is applied by assigning $\mu_1 = 1.2$~eV to the active patch (either left or right meta-network) and $\mu_2 = 0.1$~eV to the surrounding regions. At $\mu_1 = 1.2$~eV, LSPRs are excited, launching virtual SPPs that radiate into free space. Theoretical modeling based on coupled-mode analysis predicts that this asymmetric excitation results in a tilted main lobe at $23^\circ$ when the right patch is active, and at $-23^\circ$ when the left patch is activated, as confirmed by full-wave simulations in Figure~\ref{fig:fig10}(b). This configuration enables rapid, programmable beam steering without mechanical components. The MetaSwitcher achieves a measured gain and directivity of $14.6$~dBi and $15.9$~dBi, respectively, with a $3$~dB beamwidth of $22.3^\circ$ and a side lobe level of $-13.7$~dB—demonstrating efficient, focused, and reconfigurable radiation suitable for on-chip THz wireless communication systems. 
This confirms that the metasurface can steer beams without changing the operating frequency, enabling time-division beam switching and adaptive routing suitable for reconfigurable WiNoC and chiplet communication.

\subsection{Multidirectional Radiation Pattern}
\begin{figure*}
\centering         
\includegraphics[trim={0cm 4.5cm 0cm 2.5cm},clip,width=\textwidth]{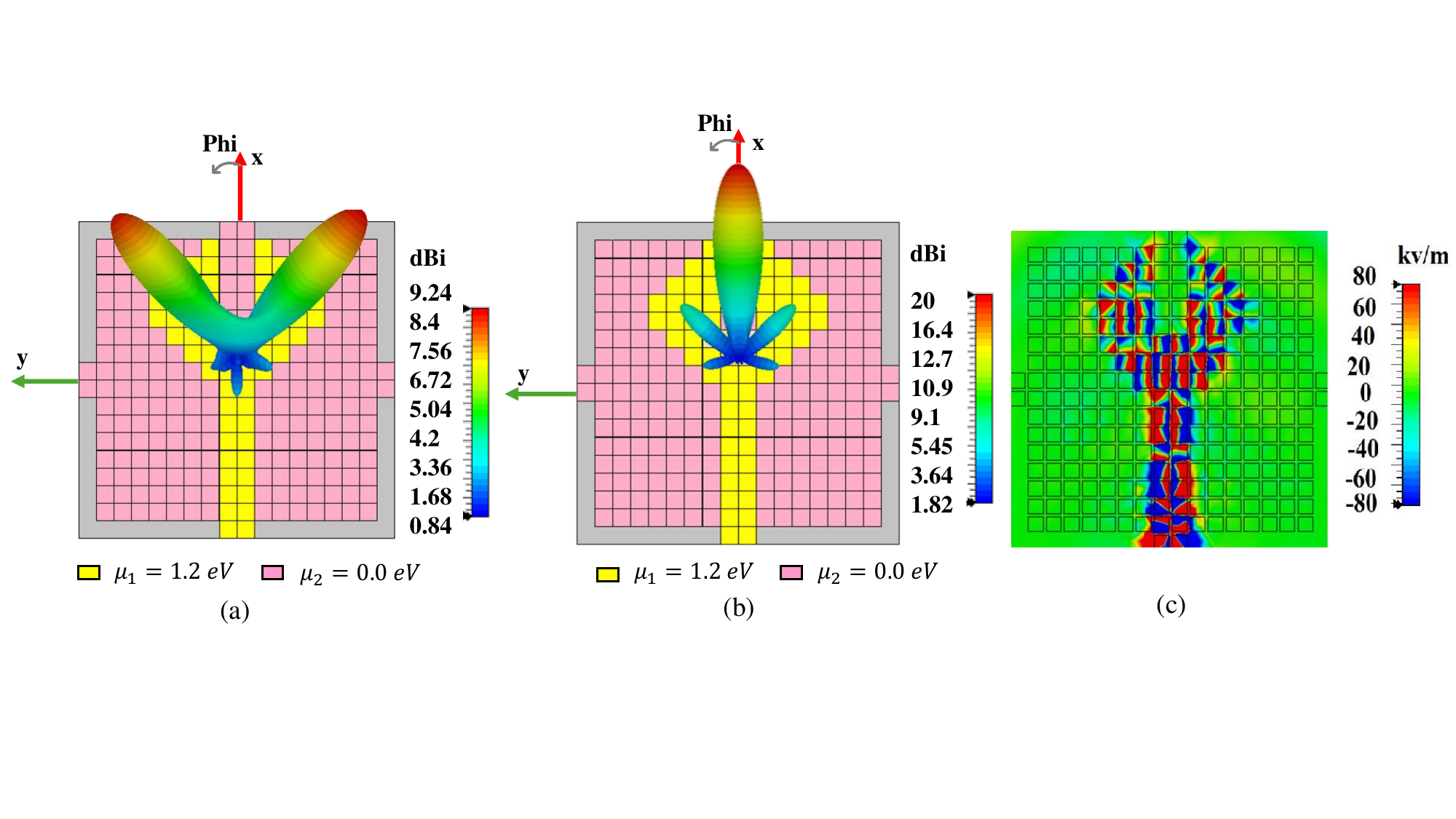}
         \label{fig:fig11}
    \caption{
Performance characteristics of the proposed \textit{Penta-MetaEmitter} antenna. 
(a) Top view of the pentagonal-shaped meta-pixel array and its 3D far-field radiation pattern at $1.65$~THz, and 
(b) at $1.7$~THz, respectively, illustrating the antenna’s ability to efficiently focus electromagnetic energy in multiple directions. 
(c) Electric field ($E$-field) distribution on the $x$–$y$ plane, demonstrating the excitation of LSPRs, which guide the design of the \textit{Penta-MetaEmitter} through the optimized selection of graphene chemical potentials for generating virtual SPPs.
}
    \label{fig:fig11}
\end{figure*}

Designing meta-pixel antennas capable of bi-directional and tri-directional radiation patterns is a key enabler for multi-user THz wireless communication systems. Bi-directional antennas, which emit energy in two opposing directions, allow a single antenna to efficiently cover users or nodes located in divergent paths—minimizing hardware complexity and maximizing spatial coverage \cite{FAKHARIAN2022168431}. Tri-directional configurations further extend this benefit by enabling signal propagation in three angular sectors, which is especially advantageous in complex layouts such as indoor environments with multiple rooms or outdoor deployments with clustered users \cite{Khodadadi_2020}. Although these patterns involve splitting the radiated power, potentially reducing transmission distance, they greatly simplify the communication architecture by eliminating the need for real-time beam steering or multiple antenna elements. This simplification is critical for intra/inter on-chip wireless connections, where fast, low-latency, and directional signal routing is essential. Moreover, directed radiation improves signal integrity at receiving elements by enhancing constructive multipath interference and minimizing unwanted reflections. Ultimately, the use of bi- and tri-directional meta-pixel antennas enhances coverage flexibility, reduces system complexity, and enables scalable network design tailored to THz communication challenges \cite{FAKHARIAN2022168431, Khodadadi_2020}.

\textbf{Penta-MetaEmitter Configuration}: Figure~\ref{fig:fig11} showcases the performance of the proposed \textit{Penta-MetaEmitter} antenna, which employs a semi-pentagonal configuration of graphene meta-pixels to achieve directional THz radiation control. As shown in Figure~\ref{fig:fig11}(a), the antenna generates a bi-directional radiation pattern at 1.65~THz with main lobes at $+41^\circ$ and $139^\circ$, achieving a peak directivity of 9.24~dBi and a $3$~dB beamwidth of $35.6^\circ$. The chemical potential of $\mu_1 = 1.2$~eV enables the excitation of THz plasmons, allowing the graphene patches to behave in a metallic regime and launch LSPRs efficiently. At a slightly higher frequency of 1.7~THz, shown in Figure~\ref{fig:fig11}(b), the antenna reconfigures into a unidirectional mode, demonstrating its frequency-dependent adaptability. The electric field distribution in the $x$–$y$ plane (Figure~\ref{fig:fig11}(c)) reveals the localized plasmonic behavior that underpins the beam shaping mechanism, emphasizing the importance of precise chemical potential tuning for activating virtual SPPs and achieving efficient EM control.
\begin{figure}
\centering
         \begin{subfigure}[b]{0.3\textwidth}
         \centering
         \includegraphics[trim={3cm 9cm 3cm 9cm},clip,width=\textwidth]{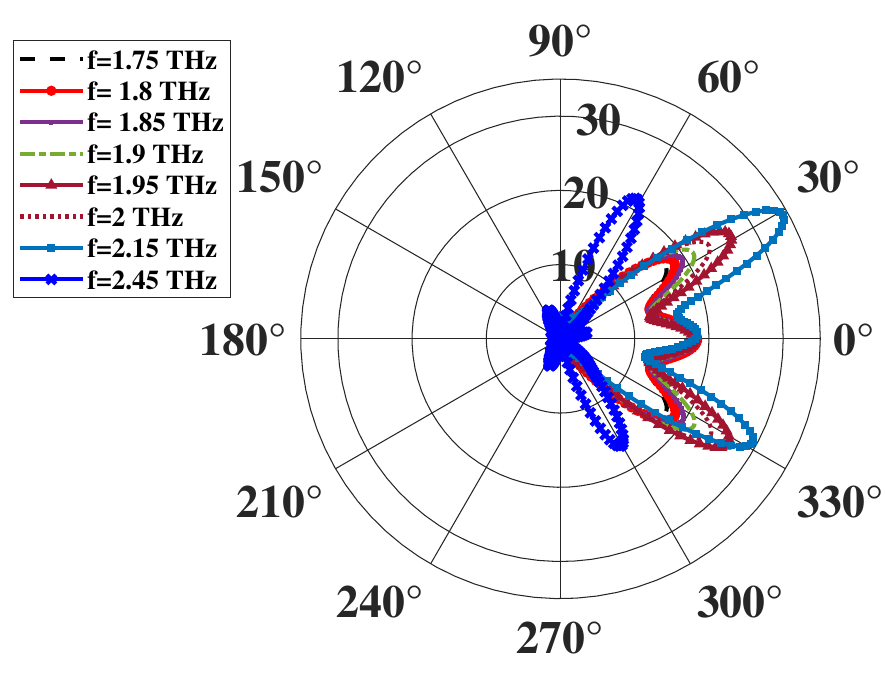}
         \caption{}
         \label{fig:fig12a}
     \end{subfigure}
              \begin{subfigure}[b]{0.4\textwidth}
         \centering
         \includegraphics[width=\textwidth]{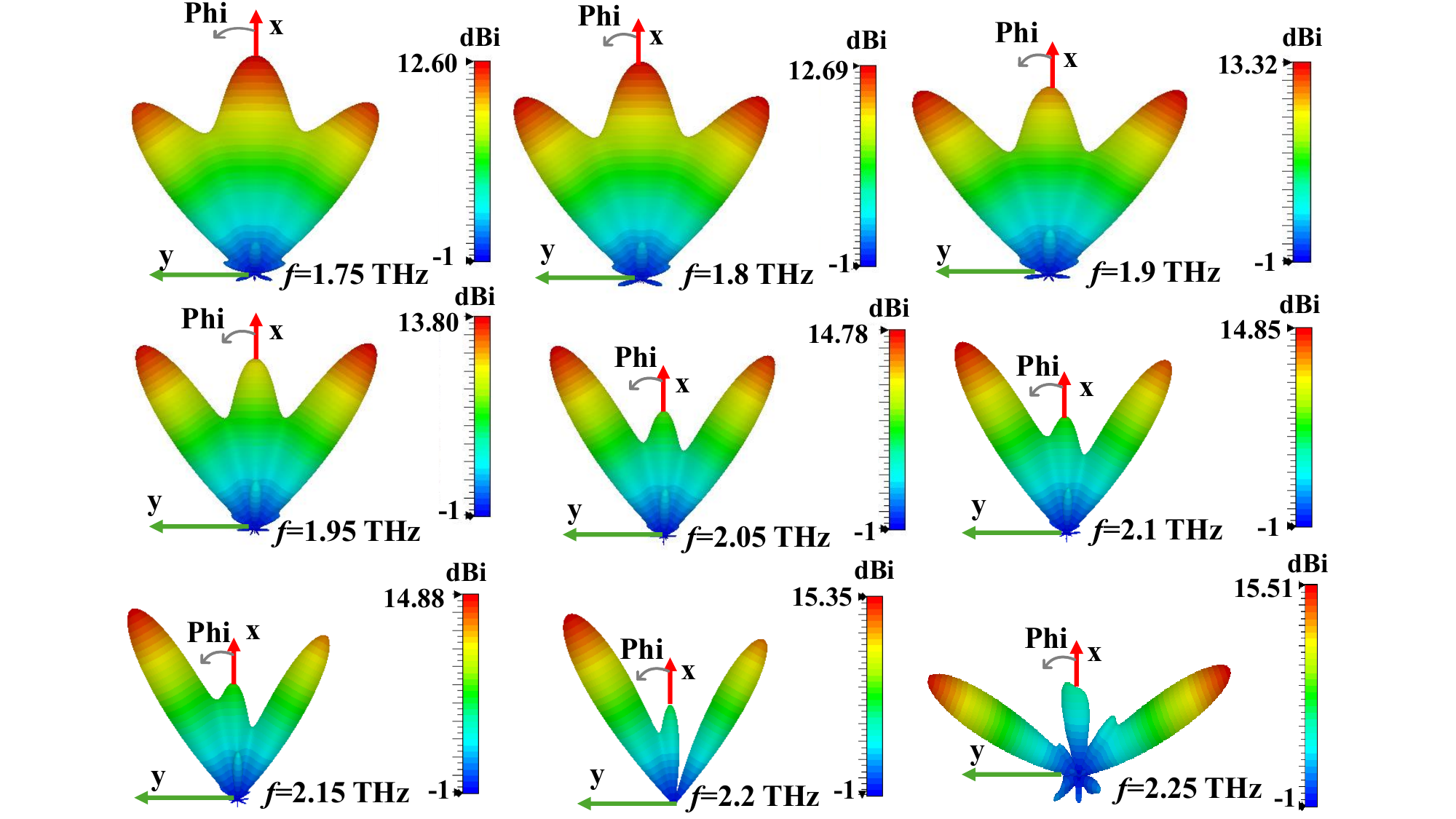}
         \caption{}
         \label{fig:fig12b}
     \end{subfigure}
    \caption{
Investigation of the radiation pattern transition in the \textit{Penta-MetaEmitter} antenna as the Tri-Gate MetaRouter. (a) Theoretical polar plot and (b) numerical far-field radiation pattern, illustrating the shift from tri-directional to bi-directional behavior with increasing frequency.}
    \label{fig:fig12}
\end{figure}

\begin{figure}
\centering
         \begin{subfigure}[b]{0.3\textwidth}
         \centering
         \includegraphics[trim={3cm 9cm 3cm 9cm},clip,width=\textwidth]{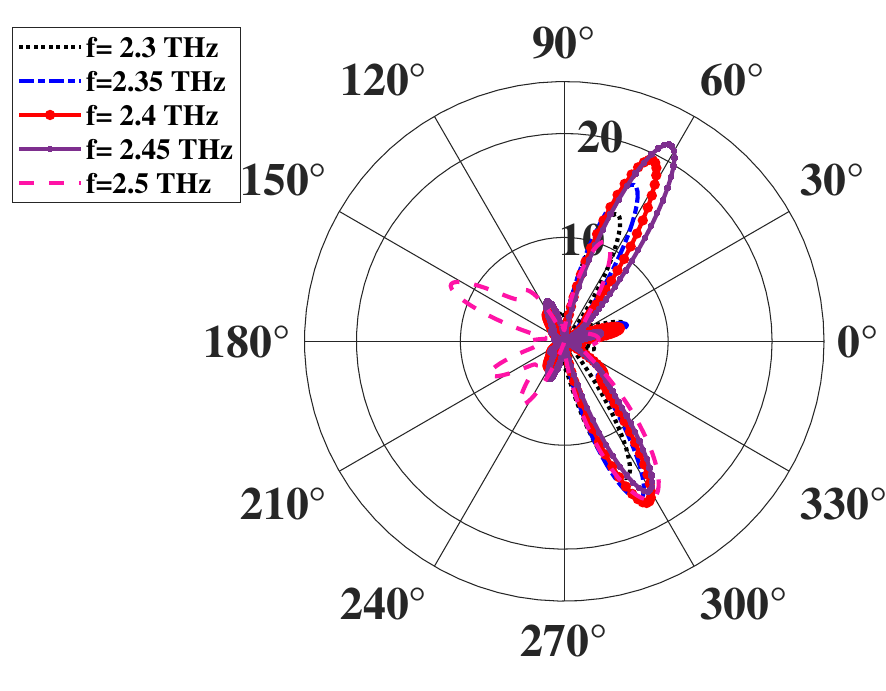}
         \caption{}
         \label{fig:fig13a}
     \end{subfigure}
              \begin{subfigure}[b]{0.3\textwidth}
         \centering
         \includegraphics[trim={4.75cm 1cm 8cm 0cm},clip,width=\textwidth]{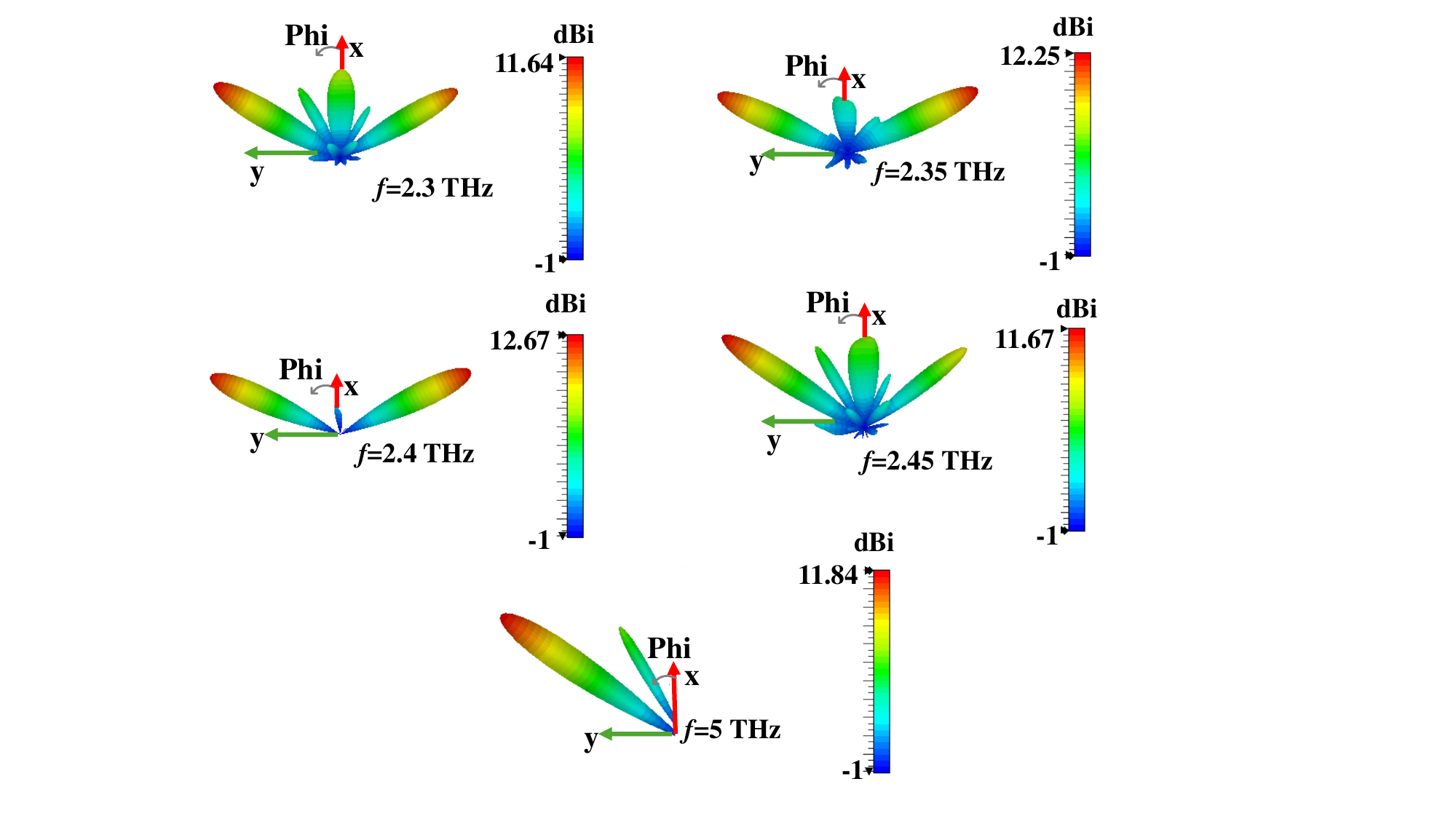}
         \caption{}
         \label{fig:fig13b}
     \end{subfigure}
    \caption{Investigation of the radiation pattern transition in the \textit{Penta-MetaEmitter} antenna as the Directional Meta-Gate. (a) Theoretical polar plot and (b) numerical far-field radiation pattern, illustrating the shift from bi-directional to unidirectional behavior with increasing frequency.}
    \label{fig:fig13}
\end{figure}

\textbf{Frequency-Driven Directional Transitions}: Figures~\ref{fig:fig12} and~\ref{fig:fig13} provide a comprehensive view of the frequency-driven beam steering behavior of the \textit{Penta-MetaEmitter}. At lower frequencies (e.g., $1.7$~THz), the radiation pattern exhibits tri-directional characteristics, with lobes directed around $\pm 60^\circ$ and $0^\circ$, as shown in the theoretical and numerical plots in Figure~\ref{fig:fig12}. As the frequency increases to $2.3$~THz, the radiation transitions to a bi-directional configuration, eliminating the central lobe and concentrating power symmetrically around $\pm 40^\circ$. This behavior reflects the redistribution of modal energy due to changing effective surface impedance and plasmonic resonance conditions within the metasurface.

Further frequency scaling from $2.3$~THz to $2.5$~THz, depicted in Figure~\ref{fig:fig13}, induces a unidirectional radiation pattern. At $2.5$~THz, most of the radiated energy is focused near the broadside direction ($\sim 5^\circ$), as observed in both theoretical and numerical results. This transition is not only directional but also improves the antenna's spatial selectivity, with directivity increasing from $11.64$~dBi to $11.84$~dBi. Such a shift is indicative of beam compression and improved impedance matching as resonance conditions align optimally at higher frequencies.
This progression—from tri- to bi- to unidirectional emission—demonstrates the metasurface’s inherent frequency scanning capability. Functionally, the \textit{Penta-MetaEmitter} operates similarly to a leaky-wave antenna, where the main beam angle is a function of operating frequency. Unlike traditional beam-steering systems that require mechanical or electrical phase control, the proposed design achieves directional control passively, by exploiting frequency-dependent phase responses and LSPR coupling.
The frequency-driven beam steering demonstrated by the \textit{Penta-MetaEmitter} has significant implications for a wide range of emerging applications in THz systems. In on-chip and chip-to-chip wireless communication architectures, such as WiNoC and chiplet-based integration, frequency-selective beam routing enables low-latency, interference-aware links by spatially directing data streams without requiring active phase shifters or mechanical elements. This passive beam steering mechanism also supports spectral beam multiplexing, where different THz frequencies are assigned to different spatial directions, enabling simultaneous multi-directional transmission within compact chip-scale platforms. Beyond chip-level systems, this approach is valuable in THz imaging, remote sensing, and radar applications, where frequency-controlled directional scanning can eliminate the need for mechanical beam steering, offering benefits in speed, reliability, and miniaturization. In spaceborne systems, including CubeSats and satellite payloads, frequency-driven metasurfaces provide a lightweight and energy-efficient means of reconfigurable beam targeting for inter-satellite links or ground communication. These characteristics make the proposed metasurface architecture a promising candidate for next-generation THz transceivers, where passive, scalable, and frequency-adaptive radiation control is essential. However, this frequency-driven steering introduces practical trade-offs. For systems where the user or receiver position is fixed, such beam shifting could lead to signal degradation across the bandwidth. This limitation is inherent in all frequency scanning systems and suggests the need for complementary control schemes (e.g., real-time reconfigurable biasing or hybrid metasurface-electronic architectures) to maintain beam alignment across a broader operating range.

\begin{figure}
\centering
         \begin{subfigure}[b]{0.2\textwidth}
         \centering
         \includegraphics[trim={10cm 3cm 11cm 1cm},clip,width=\textwidth]{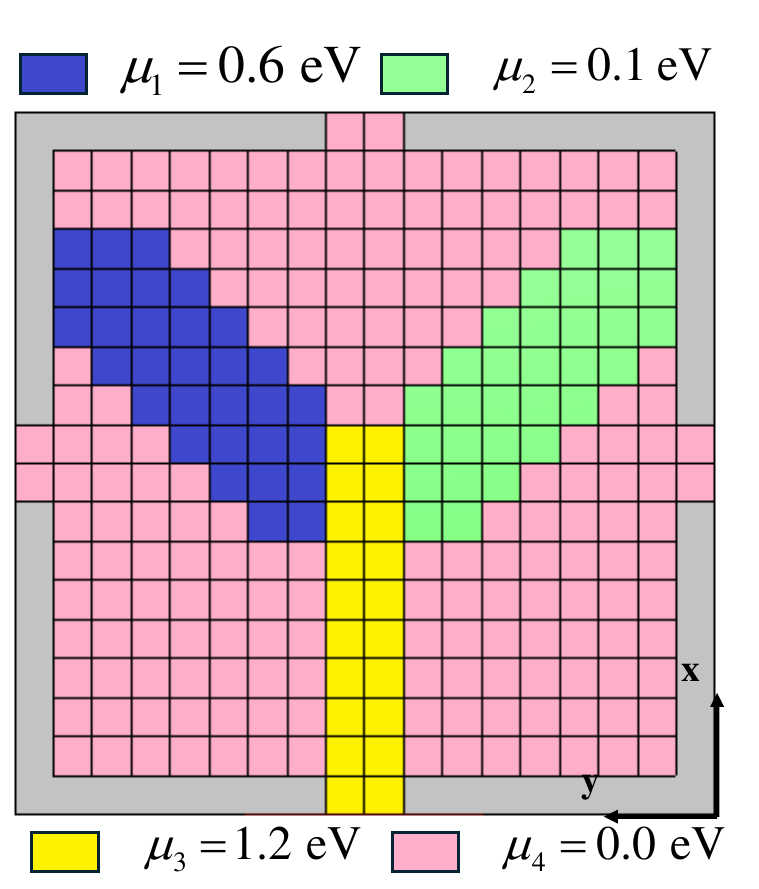}
         \caption{}
         \label{fig:fig14a}
     \end{subfigure}
              \begin{subfigure}[b]{0.28\textwidth}
         \centering
         \includegraphics[trim={3cm 9.3cm 4cm 9.3cm},clip,width=\textwidth]{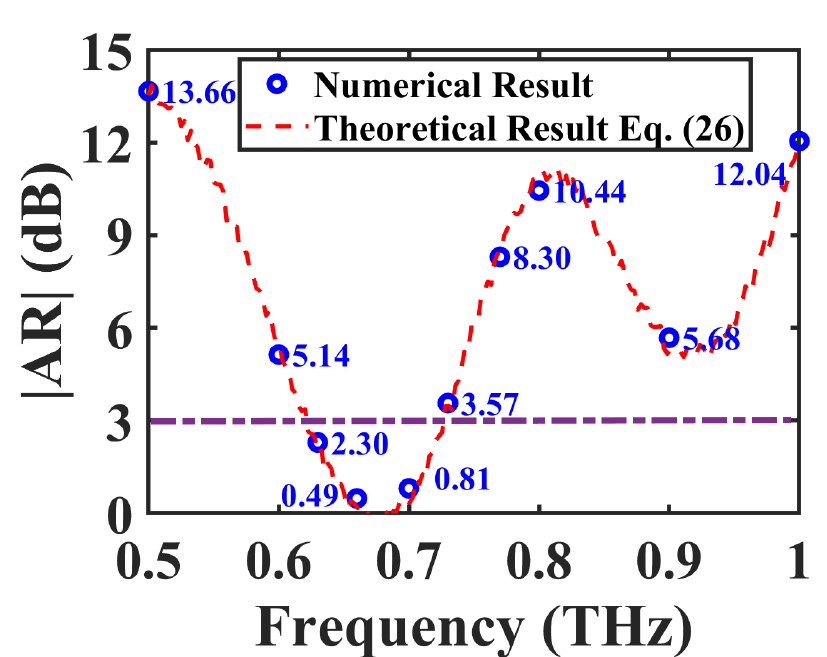}
         \caption{}
         \label{fig:fig14b}
     \end{subfigure}
       \begin{subfigure}[b]{0.4\textwidth}
         \centering
         \includegraphics[width=\textwidth]{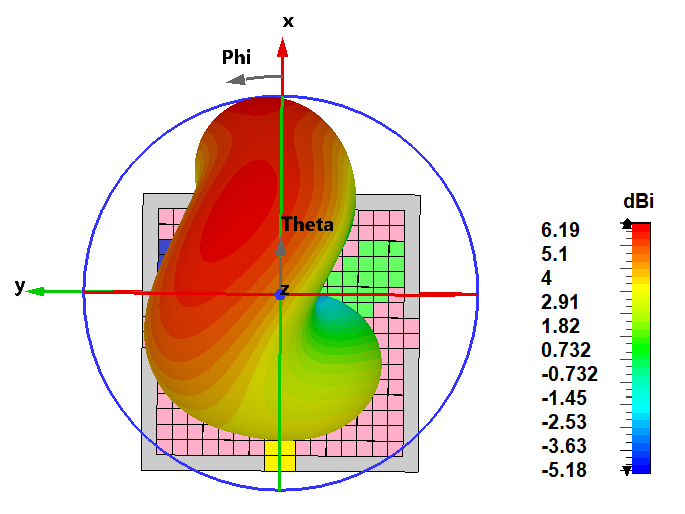}
         \caption{}
         \label{fig:fig14c}
     \end{subfigure}
    \caption{\textit{CP-MetaCore} antenna configuration and polarization performance. (a) Circularly polarized meta-pixel antenna configuration designed for THz WiNoC environments, enabling robust and orientation-independent communication. (b) Theoretical and numerical axial ratio spectra over the range of $0.65$–$0.72$~THz, demonstrating effective circular polarization (CP) with values consistently below $3$~dB, confirming the antenna’s suitability for stable CP-based wireless communication links. (c) Far-field radiation pattern at the frequency of $0.7$~THz.}
    \label{fig:fig14}
\end{figure}

\subsection{CP-MetaCore Configuration}

Circular polarization antenna plays a crucial role in WiNoC systems by enabling robust, orientation-independent communication with enhanced resilience to multipath interference. In densely integrated chip environments where reflections and unpredictable wave orientations are common, CP helps maintain signal integrity and improves transmission reliability. The proposed \textit{CP-MetaCore} configuration, Figuere~\ref{fig:fig14}(a), leverages this property to support wideband, stable THz links for dynamic and efficient on-chip wireless interconnects. The CP-MetaCore antenna demonstrates effective circular polarization within the 0.65–0.72~THz range, as indicated by the axial ratio remaining below 3~dB. Figure~\ref{fig:fig14}(b) presents both theoretical and numerical results confirming this performance. As shown in Figure~\ref{fig:fig14}(c), although its directivity ($6.19$~dBi) and total efficiency ($85\%$) are lower than other configurations, the stable axial ratio ensures reliable circularly polarized THz communication within the specified band. The twisted radiation pattern observed in Figure~\ref{fig:fig14}(c) is attributed to slight phase and amplitude imbalances between orthogonal field components, resulting in an elliptically polarized wavefront. This is consistent with the axial ratio results near the 3~dB boundary, where the polarization deviates slightly from ideal circularity. Such behavior is typical in metasurface antennas operating at the edge of their CP bandwidth, especially when asymmetric field excitation, non-uniform chemical potential distribution, or spatial variations in impedance tuning occur across the structure.

\begin{figure*}
   \centering
         \begin{subfigure}[b]{0.3\textwidth}
         \centering
        \includegraphics[width=\textwidth]{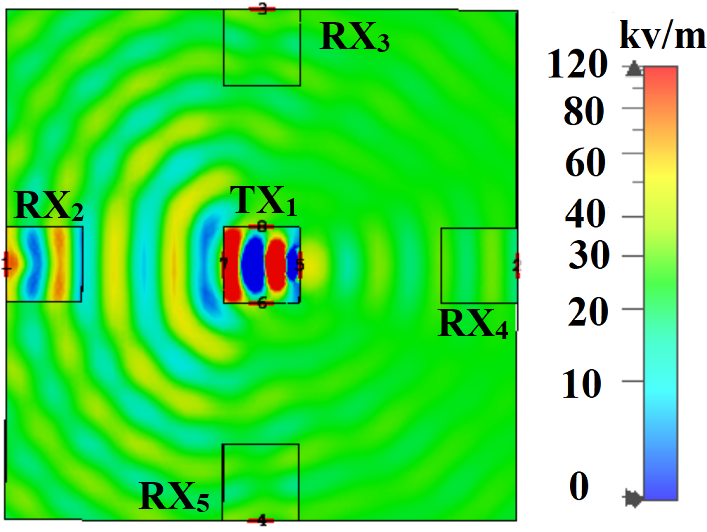}
         \caption{}
         \label{fig:fig14a}
     \end{subfigure}
              \begin{subfigure}[b]{0.3\textwidth}
         \centering
         \includegraphics[trim={3cm 9.2cm 4.2cm 9.8cm},clip,width=\textwidth]{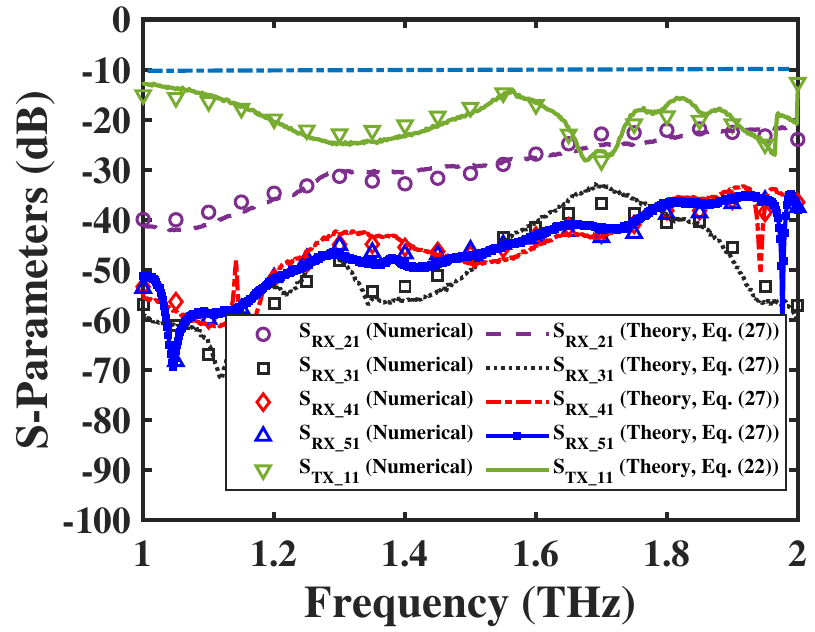}
         \caption{}
         \label{fig:fig15b}
     \end{subfigure}
     \begin{subfigure}[b]{0.7\textwidth}
         \centering
         \includegraphics[trim={0cm 6cm 0cm 4cm},clip,width=\textwidth]{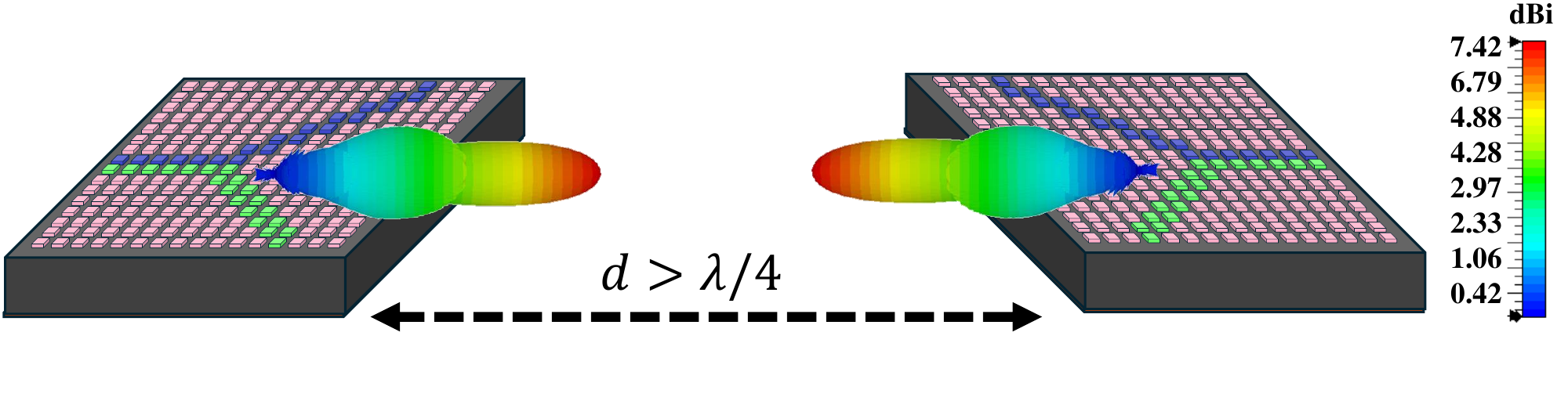}
         \caption{}
         \label{fig:fig15c}
     \end{subfigure}
    \caption{Point-to-point meta-wireless link performance. (a) Electric field distribution at $1.8$~THz, illustrating the excitation and propagation of virtual SPPs across the meta-network. (b) Comparison between theoretical and full-wave simulation results for the S-parameters: transmission ($S_{21}$) and reflection ($S_{11}$) spectra. (c) 3D far-field radiation pattern at $1.8$~THz, demonstrating directional beamforming behavior of the meta-wireless link.}
    \label{fig:fig15} 
\end{figure*}

\subsection{State-of-the-Art Comparison}

To substantiate the performance of the proposed meta-network architectures, a comprehensive comparison is presented in Table~\ref{tab:comparison}, contrasting them with recent state-of-the-art reconfigurable THz metasurface antennas. The table outlines key performance metrics, including beamwidth, gain, Maximum directivity, modulation speed, tunability, and reconfiguration type. Among the proposed designs, the \textit{Penta-MetaEmitter} achieves notably high gain with dynamic binary control, making it well-suited for compact THz beamforming. The \textit{Y-MetaRouter} enables adaptive beam splitting across three distinct paths, offering an efficient solution for local THz network routing. The \textit{Meta-Gateway Link} supports selective beam steering between distant chiplet blocks, while the \textit{Tri-Gate MetaRouter} and \textit{Directional Meta-Gate} provide multi-directional and switchable routing paths. Notably, the \textit{MetaSwitcher} and related designs demonstrate flexible re-routing and scalable multi-node coupling—critical for complex intra-chip communication topologies. Beyond dynamic beam control, the proposed metasurfaces also support circular polarization, a feature not commonly available in prior work. The binary field-driven routing mechanism allows precise modulation of local current distributions on graphene meta-pixels, enabling the formation of rotating electric field vectors essential for CP emission. This functionality introduces an additional degree of freedom for polarization diversity, enhancing robustness in complex THz chip environments.

Compared to existing platforms, such as VO$_2$-based metasurfaces which suffer from slow thermal switching~\cite{yang2022programmable, 10259207}, or PIN-diode arrays that offer limited discrete states and require bulky external circuitry~\cite{app122211780}, the proposed architectures utilize electrostatically biased graphene. This enables sub-nanosecond response speeds with fine-grained binary control. Although continuously tunable Fermi-level metasurfaces~\cite{9495239} may achieve higher peak gain, they lack inherent topological reconfigurability and are less compatible with chiplet-scale, binary-controlled communication systems. Similarly, reflectarray-type metasurfaces~\cite{tamagnone2018reflectarray} offer broad angular coverage but do not support internal path reconfiguration or virtual routing.
In contrast, the architectures introduced in this work offer digitally programmable, binary-LSPRs for THz routing, combined with optional CP generation. This enables rapid beam switching, polarization agility, and reconfigurable communication topologies—making them well-suited for ultra-compact, multi-beam THz networks operating at the chip and interposer level. The proposed metasurfaces support both spatial- and time-division multiplexing with minimal hardware overhead, establishing a robust platform for scalable and adaptive on-chip wireless communication.

\begin{table*}[htbp]
\centering
\caption{Performance comparison of proposed metasurface architectures with state-of-the-art THz antennas.}
\label{tab:comparison}
\begin{tabular}{|>{\centering\arraybackslash}p{3cm}|>{\centering\arraybackslash}p{1.8cm}|>{\centering\arraybackslash}p{1.7cm}|>{\centering\arraybackslash}p{2.2cm}|>{\centering\arraybackslash}p{2.1cm}|c|>{\centering\arraybackslash}p{2.5cm}|}
\hline
\textbf{Design} & \textbf{Beamwidth \newline (°)} & \textbf{Gain \newline(dBi)} & \textbf{Max. Directivity \newline
(dBi)} & \textbf{Mod. Speed} & \textbf{Tunability} & \textbf{Reconfig. Type} \\
\hline
\multicolumn{7}{|c|}{\textbf{This Work}} \\
\hline
\textbf{Y-MetaRouter} & 65.4 & 15.1 & 15.3 & Fast (Biasing) & Dynamic & LSPR Coupling \\
\textbf{Meta-Gateway Link} & Multi-angle & 6.69-11.84 & $\sim$15 & Fast (Biasing) & Dynamic & LSPR Coupling \\
\textbf{Penta-MetaEmitter} & 15 & 20 & 20.1 & Fast (Biasing) & Dynamic & LSPR Coupling  \\
\textbf{MetaSwitcher} & $\pm$22.3 & 14.6 & 15.9 & Fast (Biasing) & Dynamic & LSPR Coupling \\
\textbf{CP-MetaCore} & -- & 6.1 & 6.19 & Fast (Biasing) & Dynamic & LSPR Coupling \\
\textbf{Tri-Gate MetaRouter} & Tri-/Bi-Dir. Switchable & 12.60-15.51 & $\sim$15.44 & Fast (Biasing) & Dynamic & LSPR Coupling \\
\textbf{Directional Meta-Gate} & Bi-/Uni-Dir. Switchable &  11.64-12.67 & $\sim$13.98 & Fast (Biasing) & Dynamic & LSPR Coupling \\
\hline
\multicolumn{7}{|c|}{\textbf{State-of-the-art}} \\
\hline
\cite{yang2022programmable} & 20--30 & $\sim$8.0 & $\sim$8.5 & Slow (Thermal) & Limited & Thermal Phase \\
\cite{app122211780} & $\sim$25 & 10.0 & 10.5 & Moderate (ns) & Partial & Discrete Switching \\
\cite{9495239} & -- & $\sim$12 & -- & Fast ($<\mu$s) & Continuous & Fermi Tuning \\
\cite{tamagnone2018reflectarray} & 60 & -- & -- & Fast (Biasing) & Continuous & Reflectarray Bias \\
\hline

\multicolumn{7}{|c|}{\textbf{Modulation=Mod., Reconfiguration=Reconfig., Maximum=Max., Directional=Dir.}} \\
\hline
\end{tabular}
\end{table*}

\section{Point-to-Point Meta-Wireless Links}

Wireless interconnects offer a transformative approach to intra-chip data exchange by eliminating physical wiring and enabling spatially agile, reconfigurable links. In the proposed metasurface-based architecture, directional communication is established by exciting selected meta-pixel units to form a dynamic transmitter (TX), while a corresponding meta-pixel array acts as the receiver (RX). To quantify the effectiveness of the wireless link between different TX and RX meta-network configurations illustrated in Figure~\ref{fig:fig1}, we generalize the classical Friis transmission equation to support multiple communication directions. Each meta-core is equipped with four antenna ports located on the top, bottom, left, and right edges, any of which can dynamically serve as the transmitter or receiver depending on spatial routing and traffic requirements. Let the horizontal and vertical distances between a selected transmitter and receiver pair be denoted as $d_x^{(j)}$ and $d_y^{(i)}$, respectively, where $i, j \in \{1,2,3,4\}$ correspond to the antenna ports (top, bottom, left, right). The generalized Friis transmission equation then becomes:

\begin{align}
P_r^{(i,j)} &= \eta_t^{(j)} \eta_r^{(i)} G_t^{(j)}(\theta_{ij}, \phi_{ij}) G_r^{(i)}(\theta_{ij}, \phi_{ij}) \nonumber \\
&\quad \times \frac{\lambda_0^2}{\left[4\pi \sqrt{(d_x^{(j)})^2 + (d_y^{(i)})^2}\right]^2} P_{\text{emit}}^{(j)}.\label{eq:eq27}
\end{align}
where $P_r^{(i,j)}$ is the received power at receiver port $i$ from transmitter port $j$, $\eta_t^{(j)}$ and $\eta_r^{(i)}$ are the respective radiation efficiencies, and $G_t^{(j)}(\theta_{ij}, \phi_{ij})$, $G_r^{(i)}(\theta_{ij}, \phi_{ij})$ are the directional gains of the transmitting and receiving antennas in the direction $(\theta_{ij}, \phi_{ij})$, Eq.~(\ref{eq:eq24}). The term $P_{\text{emit}}^{(j)}$ denotes the power emitted in the direction of the receiver, computed from the far-field electric field $E(\theta,\phi)$ obtained via the metasurface array factor, Eq.~(\ref{eq:eq23}). The total distance $d = \sqrt{(d_x^{(j)})^2 + (d_y^{(i)})^2}$ reflects the spatial separation between the chosen antenna ports, and $\lambda_0$ is the free-space wavelength.

By considering the setup in Figure~\ref{fig:fig15}(a), the electric field distribution illustrates how the excitation of LSPRs generates virtual SPPs for on-chip wireless data transfer. The system demonstrates dynamic control of transmission paths by selectively activating ports and enabling or disabling receiver modules, such as $RX_{2}$, to achieve real-time, multi-user access and programmable multi-beam routing. To validate this directional meta-link model, the theoretical prediction of transmission spectra based on the generalized Friis equation, Eq.~(\ref{eq:eq27}), is compared with full-wave simulation results, showing strong agreement as presented in Figure~\ref{fig:fig15}(b). Finally, the 3D far-field radiation pattern at 1.8~THz, shown in Figure~\ref{fig:fig15}(c), demonstrates how the meta-pixel transceiver ($\text{TX}_1$) effectively focuses energy toward the designated receiver ($\text{RX}_2$), thereby enhancing the efficiency of the THz communication link. These results confirm the feasibility and accuracy of the proposed field-driven metasurface for dynamic, high-performance on-chip wireless communication.

\subsection{System-Level Performance Estimates}

We estimate the communication performance of the Y-MetaRouter based on its simulated characteristics at 1.8\,THz, where it exhibits a gain of 15.1\,dBi and a directivity of 15.3\,dBi. The effective operational bandwidth is approximately 1.5\,THz, and the corresponding 3\,dB angular beamwidth is 65.4°, as shown in Figure~\ref{fig:fig8}.

\textbf{Link Budget:} Assuming a transmit power of 0\,dBm and using the Friis equation [Eq.~(\ref{eq:eq27})], the free-space path loss over a 1.5\,mm chiplet link at 1.8\,THz is approximately 62.6\,dB, as consistent with the geometry in Figure~\ref{fig:fig15}. An additional 1\,dB impedance mismatch loss is included due to graphene–metal transitions~\cite{pham2017broadband}, along with a 1.1\,dB packaging and dielectric loss based on silicon interposer measurements at sub-THz frequencies~\cite{imec2023interposer, 10.1063/5.0211061}. This yields a total link attenuation of approximately 64.7\,dB. Given transmit and receive gains of 15.1\,dBi, the resulting received power is estimated as –34.5\,dBm.

\textbf{Data Rate:} The theoretical maximum data rate is estimated using the Shannon–Hartley capacity formula \cite{6005345}:

\begin{equation}
R = B \cdot \log_2(1 + \text{SNR}),
\label{eq:shannon}
\end{equation}
where \( B \) is the system bandwidth and SNR is the signal-to-noise ratio. For the given link budget and thermal noise floor, the SNR is estimated at approximately 41.8\,dB, which results in a theoretical capacity of about 10.5\,Tbps. In practice, spectral efficiency is limited by modulation order and coding. A 16-QAM modulation scheme is a reasonable compromise between complexity and performance for short-range THz links~\cite{6005345}, offering 4 bits/symbol. Assuming a conservative 20\% forward error correction (FEC) overhead~\cite{11016733}, the net achievable throughput is estimated in the range of 3–5\,Tbps.

\textbf{Beam Alignment Tolerance:} For a 3\,dB beamwidth of 65.4°, the corresponding half-power angle is 32.7°. Given a communication distance of 1.5\,mm, the lateral misalignment tolerance can be approximated as $\Delta x = 1.5 \cdot \tan(32.7^\circ) \approx 0.96\,\text{mm},$ which represents the maximum permissible lateral offset before incurring more than 3\,dB of link degradation. This tolerance range is compatible with current chiplet alignment capabilities in silicon interposer and advanced packaging platforms~\cite{Jiang_2007}.

\textbf{Control Power and Switching Energy:} Each graphene patch is controlled via electrostatic gating, allowing modulation of its local conductivity through an externally applied voltage. For a typical gate capacitance of 10\,fF and a voltage swing of 2.5\,V, the switching energy per patch is estimated using:

\begin{equation}
E = \frac{1}{2} C V^2,
\label{eq:switching_energy}
\end{equation}
where \( C \) is the gate capacitance and \( V \) is the applied bias voltage. This yields an energy of approximately 31.25\,fJ per pixel~\cite{ju2011graphene}. Given that the system consists of a \(16 \times 16\) array of monolayer graphene patches, the total energy required for full-array reconfiguration is approximately 8\,nJ. This ultra-low energy budget highlights the suitability of graphene metasurfaces for fast, energy-efficient dynamic beam steering.

\textbf{Latency:} The switching speed is limited by bias settling and control logic delay, both expected to be below 1\,ns. Total reconfiguration latency is estimated at 1–2\,ns \cite{phare2015graphene}.

\section{Conclusion}
This paper introduces a novel field-driven LSPR meta-network framework for dynamic and reconfigurable plasmonic THz on-chip communication. By leveraging tunable graphene meta-pixels, we proposed a set of antenna configurations capable of real-time beam steering, adaptive polarization control, and multi-user access. The \textit{Binary Field-Driven Meta-Routing Method} enables programmable excitation of LSPRs, forming virtual SPPs pathways that facilitate efficient wireless data transmission across chip domains. A generalized coupled-mode theoretical model was introduced and validated against full-wave simulations, showing strong agreement in both field distribution and radiation characteristics. Directional gain, polarization purity, and S-parameters confirm the practical potential of the proposed approach.  Furthermore, the proposed architecture was validated through a point-to-point meta-wireless link scenario, in which its ability to dynamically activate specific transmitter and receiver pairs with high directionality and communication efficiency was demonstrated.
The system's performance was analyzed both theoretically, using the generalized Friis equation adapted to the metasurface framework, and numerically, through full-wave electromagnetic simulations. The strong agreement between theoretical predictions and simulation results confirms the feasibility of meta-network THz links for dense, multi-user chip environments. 

System-level performance estimates confirm the viability of this approach, with link budgets showing received power levels above practical thresholds, theoretical data rates reaching 10.5~Tbps, and achievable throughput in the 3–5~Tbps range using 16-QAM modulation. Beam alignment tolerance and switching energy analysis highlight the robustness and energy-efficiency of the proposed system, with reconfiguration energy below 10~nJ and latency on the order of 1–2~ns. These metrics establish a compelling case for metasurface-enabled THz wireless links in dense chiplet environments.Looking ahead, this work lays the foundation for practical deployment in advanced chiplet-based computing platforms, WiNoC architectures, and future 6G microelectronic systems. Future efforts will focus on scalable integration, CMOS compatibility, and intelligent control mechanisms to support real-time, energy-efficient, and adaptive interconnects in high-performance and AI-driven computing environments.

\section*{Acknowledgments}
We would like to express our deepest gratitude to Dr. Seyyed Mohammad Mahdi Moshiri for his invaluable, insightful feedback and continuous support throughout the course of this work.




\newpage

\begin{IEEEbiography}[{\includegraphics[width=1in,height=1.25in,clip,keepaspectratio]{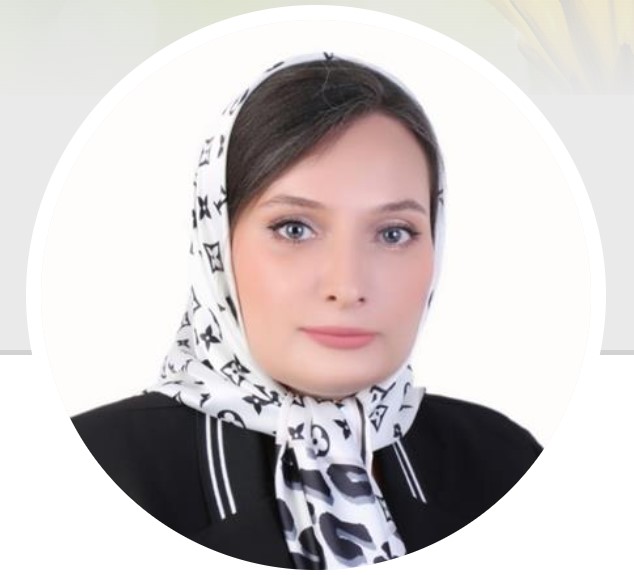}}]{ Maryam Khodadadi} (Member, IEEE),was born in Tehran, Iran in 1988. She earned her B.Sc. and M.Sc. degrees in computer engineering and telecommunication engineering from K. N. Toosi University of Technology, Tehran, Iran, in 2011 and 2015, respectively. She later obtained her Ph.D. in telecommunication engineering from Shiraz University of Technology, Shiraz, Iran, in 2020. From 2020 to 2022, she conducted postdoctoral research on controllable hybrid plasmonic integrated circuits with the Shiraz University of Technology. During this period, she received a research fellowship from the Iran National Science Foundation (INSF). Since 2023, she has held a postdoctoral research associate position at the Institute for Communication Systems (ICS) at the University of Surrey, UK, a home to the 5G and 6G Innovation Centres (5GIC and 6GIC). 
Maryam Khodadadi has been awarded the prestigious Marie Curie Fellowship, supported by EPSRC, in 2024. She has been a member of the IEEE Educational Activities Committee since 2021 and also serves as the Secretary of the Steering Committee for the Electromagnetics and Photonics chapter of the IEEE Iran Section. She became a professional member of IEEE in 2022. 
She has conducted extensive research in Reconfigurable Intelligent Surfaces (RIS) and photonic topological insulators (PTIs). Her diverse research interests include hybrid plasmonic nano-antennas, plasmonic devices as logic gates, sensors, metamaterials, absorbers, and nanostructure modeling and analysis. 
\end{IEEEbiography}

\begin{IEEEbiography}[{\includegraphics[width=1in,height=1.25in,clip,keepaspectratio]{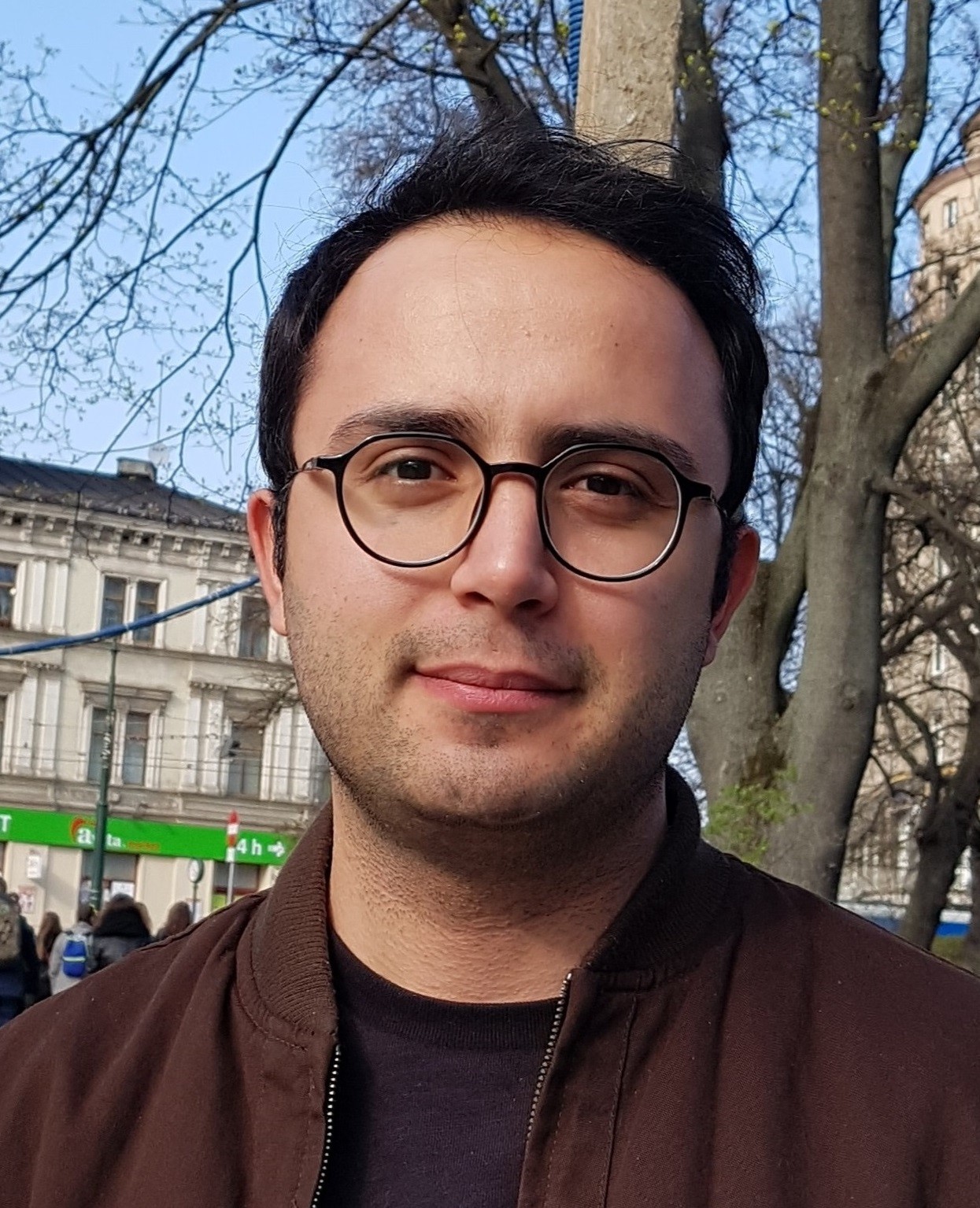}}]{Hamidreza Taghvaee} (Member, IEEE) obtained his PhD in Computer Architecture from the Polytechnic University of Catalonia, Spain, with Cum Laude honours. His doctoral research garnered significant recognition, including a €7k grant from the European Commission and €7k from the Academy of Finland for a visiting researcher stint at the Meta Group, Aalto University. His exceptional contributions during his PhD were further acknowledged with a special doctoral award for outstanding theses by the Universitat Politècnica de Catalunya in 2021. His doctoral research involved significant participation in the VISORSURF and WiPLASH FET-OPEN projects at N3Cat, Barcelona Tech. Progressing into postdoctoral research, Taghvaee contributed to the RISE-6G FET-OPEN and OBLICUE EPSRC projects at the George Green Institute, University of Nottingham. Currently, he holds a senior postdoctoral researcher position at the Institute for Communication Systems at the University of Surrey, home to the innovative 5G and 6G Innovation Centres. In this capacity, he leads the Flexi-DAS, CORE, and SCONDA projects, funded by DISIT. Moreover, he has successfully secured funding exceeding £200k from Innovate UK and over £750k from the Joint Lab initiative. Moreover, he plays an active role in shaping the future of telecommunications technology as a member of the Industry Specification Group (ISG) on Reconfigurable Intelligent Surfaces within the European Telecommunications Standards Institute (ETSI), contributing to the development of standards that will underpin the next generation of communication systems.
\end{IEEEbiography}

\begin{IEEEbiography}[{\includegraphics[width=1in,height=1.25in,clip,keepaspectratio]{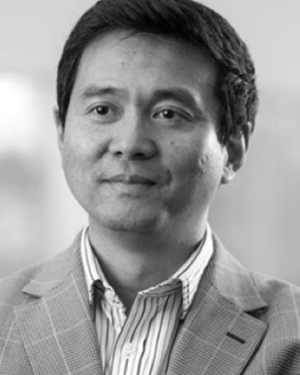}}]{Pei Xiao} (Senior Member, IEEE) is a Professor
in Wireless Communications in the Institute for
Communication Systems (ICS) at University of
Surrey. He is currently the technical manager
of 5GIC/6GIC, leading the research team
in the new physical layer work area, and
coordinating/supervising research activities
across all the work areas (https://www.
surrey.ac.uk/institute-communication-systems/
5g-6g-innovation-centre). Prior to this, he worked
at Newcastle University and Queen’s University
Belfast. He also held positions at Nokia Networks in Finland. He has published extensively in the fields of communication theory, RF and antenna
design, signal processing for wireless communications, and is an inventor on
over 15 recent patents addressing bottleneck problems in 5G/6G systems.
\end{IEEEbiography}

\begin{IEEEbiography}[{\includegraphics[width=1in,height=1.25in,clip,keepaspectratio]{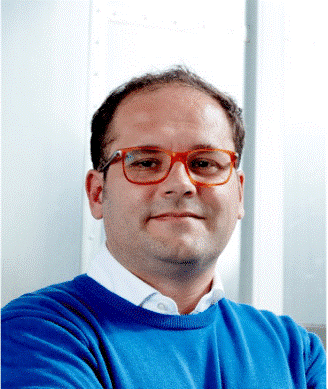}}]{Gabriele Gradoni} (Senior Member, IEEE) earned their Ph.D. in electromagnetics from Università Politecnica delle Marche, Ancona, Italy, in 2010. They further enriched their academic journey as a Visiting Researcher with the Time, Quantum, and Electromagnetics Team at the National Physical Laboratory, Teddington, U.K., in 2008. The period from 2010 to 2013 saw them contributing as a Research Associate at the Institute for Research in Electronics and Applied Physics, University of Maryland, College Park, MD, USA. From 2013 to 2016, they advanced their career as a Research Fellow at the School of Mathematical Sciences, University of Nottingham, U.K., culminating in their appointment as a Full Professor of Applied Mathematics and Electromagnetics Engineering in 2022. As of May 2023, they have been serving as Full Professor and Chair of Wireless and Satellite Communications at the 6G Innovation Centre, Institute for Communication Systems, University of Surrey, Guildford, U.K. Their role as a Royal Society Industry Fellow commenced in 2020 at British Telecom, U.K. Since December 2022, they have held positions as a Visiting Fellow at the Department of Computer Science and Technology, University of Cambridge, U.K., and as an Adjunct Professor at the Department of Electrical and Computer Engineering, University of Illinois at Urbana–Champaign, USA. Their research spans probabilistic and asymptotic methods for wave propagation in complex systems, metasurface modelling, wave chaos, and quantum computational electromagnetics. These areas of interest have significant implications for electromagnetic compatibility and the evolution of modern wireless communication systems. The individual is an esteemed member of both the IEEE and the Italian Electromagnetics Society. Their work has been recognized with several accolades, including the URSI Commission B. Young Scientist Award in 2010 and 2016, the Italian Electromagnetics Society Gaetano Latmiral Prize in 2015, and an Honorable Mention for the IEEE TEMC Richard B. Schulz Transactions Prize Paper Award in 2020. Furthermore, they have been honoured with multiple Best Paper awards at international forums, notably receiving the Best Electromagnetics Award at EuCAP 2022. Between 2014 and 2021, they represented early career professionals as the URSI Commission E Early Career Representative, highlighting their dedication to advancing the field of electromagnetic.
\end{IEEEbiography}

\begin{IEEEbiography}
[{\includegraphics[width=1in,height=1.25in,clip,keepaspectratio]{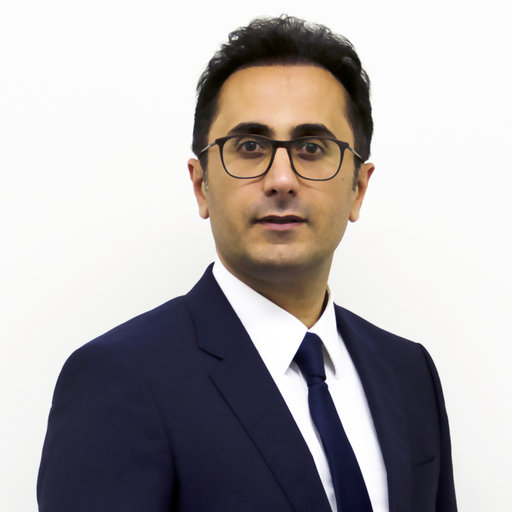}}]{Mohsen Khalily}(S'09--M'12--SM'18) is currently a Senior Lecturer in antennas and propagation and the Head of the Surface Electromagnetics Laboratory (SEMLAB) at the Institute for Communication Systems. He has published four book chapters and almost 200 academic articles in international peer-reviewed journals and conference proceedings and has been the Principal Investigator on research grants totalling in excess of £4 million in the field of surface electromagnetics. He is the leading rapporteur for the work item on implementation and practical considerations of the RIS in the dedicated Industry Specification Group within the  ETSI. His research interests include surface electromagnetics, antennas and propagation. He is a fellow of the U.K. Higher Education Academy and serves as an Associate Editor for IEEE Antennas and Wireless Propagation Letters, and Scientific Reports (Nature-Index)

\end{IEEEbiography}

\vspace{11pt}


\vfill

\end{document}